\documentclass[preprintnumbers,aps,prd,longbibliography,nofootinbib,nobibnotes,amsmath,amssymb]{revtex4}
\usepackage{graphicx,color,natbib}
\usepackage{amsmath,amssymb,amsfonts}
\usepackage{epsfig,epsf}
\usepackage{dcolumn}
\usepackage{float}
\usepackage{bm}
\usepackage{subfig}
\usepackage{caption}
\input amssym.def
\input amssym.tex

\newcommand{\nc}{\newcommand}
\nc{\no}{\nonumber}
\makeatletter

\usepackage[colorlinks=true, linkcolor=blue, bookmarks=true]{hyperref}

\begin{document}
\title{Chaos Near to the Critical Point: Butterfly Effect and Pole-Skipping}
\author{B. Amrahi\footnote{$\rm{b}_{-}$amrahi@ipm.ir}}
\affiliation{IPM, School of Particles and Accelerators, P.O. Box 19395-5531, Tehran, Iran}
\author{M. Asadi\footnote{$\rm{m}_{-}$asadi@ipm.ir}}
\affiliation{IPM, School of Particles and Accelerators, P.O. Box 19395-5531, Tehran, Iran}
\author{F. Taghinavaz\footnote{ftaghinavaz@ipm.ir}}
\affiliation{IPM, School of Particles and Accelerators, P.O. Box 19395-5531, Tehran, Iran}

\begin{abstract}
We study the butterfly effect and pole-skipping phenomenon for the 1RCBH model which enjoys a critical point in its phase diagram. Using the holographic idea, we compute the butterfly velocity and interestingly find that this velocity can probe the critical behavior of this model. We  calculate the dynamical exponent of this quantity near the critical point and find a perfect agreement with the value of the other quantity's dynamical exponent near this critical point. 
We also find that at chaos point, the phenomenon of pole-skipping appears which is a sign of a multivalued retarded correlation function. We briefly address the butterfly velocity and pole-skipping for the AdS-RN black hole solution which on its boundary a strongly charged coupled field theory lives. For both of these models, we find $v_B^2 \geq c_s^2$ at each point of parameter space where $c_s$ is the speed of sound wave propagation. 	
\end{abstract}

\maketitle

\tableofcontents

\section{Introduction}
So far, many approaches have been introduced for the concept of quantum chaos ranging from semiclassical methods to random matrix theory \cite{Bohigas:1983er}. Quantum chaos is a general characterization of strongly-interacting systems whose new realization   has been recently provided by the strongly-coupled many-body systems and, surprisingly, by the black hole physics. The latter has come from the gauge/gravity duality which relates a $d$-dimensional quantum field theory (QFT) to some ($d+1$)-dimensions classical gravitational theory in the bulk  \cite{Maldacena:1997re, Gubser:1998bc, Witten:1998qj}. Although there is no complete and precise definition of quantum chaos, studying how chaotic dynamics describe  quantum information has always been an intriguing topic and has attracted a lot of attention in the literature.  

 A distinct feature of quantum chaos is the butterfly effect  which is a very common phenomenon in many-body systems, explaining thermal behavior among other things. In the context of quantum mechanics, this phenomenon can be characterized using the commutator $\langle[W(t), V (0)]^2\rangle$ between two generic Hermitian operators $V$ and $W$ at time $t$. This quantity measures how much an early perturbation $V(0)$ affects the later measurements of $W(t)$ or, in another word, how sensitive the system is to an initial perturbation created by acting with $V(0)$. The strength of such measurement  is encoded in the following double commutator 
 \begin{align}\label{c(t)}
C(t)= - \langle [W(t),V(0)]^{2}\rangle _{\beta},
 \end{align}
 where the expectation value has been taken in a thermal state with  $\beta = 1/T$. If we assume that $V$ and $W$ are Hermitian and unitary operators, then the double commutator will be written 
 \begin{align}
 C(t)= 2 - 2 \langle W(t) V(0) W(t) V(0) \rangle _{\beta},
  \end{align}
 where the second part, $\langle W(t) V(0) W(t) V(0) \rangle _{\beta}$,  is called Out of Time Ordered Correlation (OTOC) and measures the degree of non-commutativity between $W(t)$ and $V(0)$. It also contains all the information of $ C(t)$. OTOC was first introduced by Larkin and Ovchinnikov to quantify the regime of validity of quasi-classical methods in the theory of superconductivity \cite{Larkin}. More recently, the definition of quantum chaos based on  OTOC became the focus of much research due to its applicability to many-body quantum systems. This kind of correlator, in the dual gravitational description, is related to  a high-energy collision that takes place close to the black hole horizon  \cite{Shenker:2013pqa,Shenker:2013yza,Shenker:2014cwa,Roberts:2014ifa,Roberts:2014isa,A. Kitaev1,A. Kitaev2,Ahn:2019rnq,Jensen:2016pah,Alishahiha:2016cjk,Wang:2022mcq}. Furthermore, there exist experimental proposals for measuring OTOC as a measure of chaos in quantum systems which is found in \cite{Swingle:2016var,Zhu:2016uws,Yao:2016ayk,Li:2016xhw}.
 There have also been various efforts to connect OTOC and  transport and hydrodynamic behavior which we will refer the reader to see \cite{Blake:2016jnn,Blake:2016wvh,Blake:2016sud,Blake:2017qgd,Grozdanov:2017ajz,Blake:2017ris,Grozdanov:2018atb,Lucas:2017ibu,Hartman:2017hhp,Haehl:2018izb,Hartman:2017hhp1,Gu:2016oyy,Davison:2016ngz,Patel:2016wdy,Baggioli:2018afg,Jeong:2021zhz,Kovtun:2005ev}.

 For simple operators in large $N$ systems, there exists  exponential growth which is characterized by the Lyapunov exponent $\lambda_{L}$ \cite{A. Kitaev1,A. Kitaev2}
  \begin{align}
 C(t)\propto \frac{1}{N^{2}} e^{\lambda_{L}t},
  \end{align}
 plus higher orders terms in $N^{-1}$ that causes $C(t)$ eventually saturate. There is the time scale where $C(t)$ becomes of order $ \textsc{O(1)}$ called scrambling time $t_{*}$ which controls how fast the chaotic system scrambles information. There is another time scale the dissipation time $t_d$ which characterizes the exponential decay of two-point functions, e.g., $\langle V(0)W(t)\rangle\sim e^{-\frac{t}{t_d}}$ and also controls the late time behavior of $ C(t)$.  In quantum field theories when we also separate the operators $V$ and $W$  in space, then at large distances \eqref{c(t)} will  generalize to 
  \begin{align}\label{c(t,x)}
C(t,\vec{x})= - \langle [W(t,\vec{x}),V(0,0)]^{2}\rangle _{\beta}.
 \end{align}
The butterfly effect is naturally characterized in terms of such a commutator, which expresses the dependence
of later measurements of distant operators $W(t,\vec{x})$ on an earlier perturbation $V(0,0)$. Note that the above expression is generically divergent and hence it  requires regularization by adding imaginary times to the time arguments of the operators $V$ and $W$, for example see \cite{Yoon:2019cql}. The interesting point is that \eqref{c(t,x)}, for a large class of models such as spin chain, higher dimensional Sachdev, Ye and Kitaev (SYK) model, and conformal field theories (CFT's), is roughly given by 
 \begin{align}
C(t,\vec{x})\propto  exp{\big[\lambda_{L}\big(t-t_{_*}-\frac{\vert x \vert}{v_B}\big)\big]},
  \end{align}
 where $v_B$ is the so-called butterfly velocity which describes the speed at which the  information about $V(0,0)$ will spread among the other degrees of freedom of the system in space. Accordingly, $v_B$  constraining spatial chaos spreading inside the butterfly effect cone (or effective lightcone) where for $t-t_{_*}\geq \frac{\vert x \vert}{v_B}$ we get $C(t,\vec{x})\sim \textsc{O(1)}$, while for $t-t_{_*}\leq \frac{\vert x \vert}{v_B}$ we have $C(t,\vec{x})\ll 1$.  Another point is that the Lyapunov exponent $\lambda_{L}$ is  upper bounded in terms of the Hawking temperature for a generic quantum system by  \cite{Maldacena:2015waa}
 \begin{align}
 \lambda_L \leq \frac{2\pi}{\beta },
 \end{align}
 where the bound is saturated by Einstein gravity, nature’s fastest scrambler \cite{Sekino:2008he}, and also by a variety of systems including  two-dimensional CFT's  in the large central charge limit, and strongly coupled SYK models.
 
In addition to OTOCs, recent developments have indicated that there exists a sharp manifestation  between the  retarded Green’s function of energy density and chaotic properties of many-body thermal systems, referred to as pole-skipping. This phenomenon refers to some special properties of equations of motion for linearized perturbations in spin-zero sector. First at the chaos point in complex ($\omega - k$) plane, $\omega = i \lambda_{L}$ and $k_\star = i k_0$ the $vv$ component of Einstein equation near the horizon becomes trivial and reduces one of the independent equations. Therefore, we expect a class of solutions to exist near the chaos point mixed of perturbations and the corresponding poles skip in lines with slope given by the initial values of perturbations \cite{Blake:2018leo}. Second, at the chaos point by appropriate choice of invariant fluctuations one can show that no singular solution exists near the horizon which is another sign of having multi-solutions for Einstein equations. The pole-skipping is special in the sense that chaotic behaviors are manifested at the level of linearized orders. This phenomenon was  first seen in the holographic context in \cite{Grozdanov:2017ajz} where they have shown that it  is universal for general thermal systems dual to Einstein gravity coupled to matter.  
The same pole-skipping phenomenon has also been observed in effective field theory (EFT) in two-point functions of energy density and flux \cite{Blake:2017ris} and it  holds in the context of chaotic two-dimensional CFTs with large central charge  limit \cite{Haehl:2018izb}. Moreover, the pole-skipping can be explored by the near horizon solutions of scalar field perturbations  \cite{Grozdanov:2019uhi, Grozdanov:2023txs, Blake:2019otz}. It was confirmed that for some models at special points in the complex ($\omega - k$) plane, namely at the Matsubara points $\omega = \omega_n = - 2 i \pi T n\,\, (n\in \mathbb{Z})$,  the differential equations can not be solved iteratively. This is because  the factor $(i \omega - 2 n \pi T)$ comes in front of coefficients which  leads to appear free parameters in spectrum of solutions. This ambiguity reflects the pole-skipping for  retarded Green's functions of operators living on the boundary and many lines can reach the point $\omega = \omega_n$ with different slopes.

In this paper, our goal is to study the butterfly effect and pole-skipping phenomenon for the  1RCBH model.  It is an analytical top-down string theory construction and is  obtained from 5-dimensional maximally supersymmetric gauged supergravity and is holographically dual to a 4-dimensional strongly coupled N = 4 SYM theory at finite chemical potential under a U(1) subgroup of the global SU(4) symmetry of R-charges \cite{Gubser:1998jb, Cvetic:1999rb, Cvetic:1999ne}. This model bears  a second-order phase transition at a certain point of its parameter space. As an extra example we do study pole-skipping phenomenon in AdS-RN (AdS-Reisner-Nordstrom) black hole as well. We review these models in detail in section \ref{sec2} and section \ref{sec4}, respectively. 

The organization of this paper is as follows. In section \ref{sec2} we focus on the butterfly effect which characterizes chaos by parameters such as butterfly velocity and Lyapunov exponent and study holographically these parameters in the mentioned model. Interestingly, we find that butterfly velocity can probe the critical point of corresponding strongly coupled field theory. Motivated by this, in section \ref{sec3}  the dynamical exponent of this velocity near the critical point has been calculated  and the perfect agreement with the ones reported in the literature has been found. Additionally, we examine the pole-skipping for higher Matsubara frequencies in differential equation of scalar field perturbations. We observe that at these points many unknown coefficients exist in the scalar field series which show multivalued retarded correlation function similar to the previous findings. While, at chaos point $\omega = i \lambda_{L}$ and $k_\star = i k_0$ we can not catch the property $\mbox{det} \mathcal{M}^n(\omega, k^2)=0$ where $\mathcal{M}^n(\omega, k^2)$ is the matrix of scalar field perturbations equations. It may be because the presence of charge  spoils this feature. Equations of the gauge field perturbations do not show such property. In section \ref{sec4} we study the pole-skipping phenomenon in both 1RCBH  and AdS-RN models. Indeed, we consider  a special set of fluctuations in the spin 0 sector and derive the linearized equations for these perturbations. We observe that the $v-v$ component of Einstein’s equations  near the horizon at the chaos point  becomes equivalent to the shock wave equation and we can read off the information about the chaos.  Furthermore, this component of the Einstein equations automatically satisfied at  the chaos point and one can deduce that there exists one extra regular mode emerging at this point and as a result leads to the pole-skipping in the retarded energy density correlation function.  Due to the complicated nature of master equations in the 1RCBH model, we supply MATHEMATICA files containing the linearized equations in the spin zero sector  separately along with this paper. In section \ref{sec5}, we conclude and briefly review  our results. To give more details, Appendix \ref{Appendix1} is devoted to more accurate calculations of invariant fluctuations in the Eddington-Finkelstein coordinates.  Also, in Appendix \ref{Appendix2} we provide the coefficients for  the linearized equation  of section \ref{sec4}.
\section{Butterfly effect in 1RCBH model }\label{sec2}
We would like to study the butterfly effect in a strongly coupled field theory including a critical point, using the framework of holography. To do so,  we first review the 1RCBH model.
The bulk theory is a top-down string theory construction which is a consistent truncation of the super-gravity on AdS$_5 \times$S$^5$ geometry in which one scalar field and one gauge field are coupled to the Einstein gravity. The thermal solution is an asymptotically AdS black brane geometry with a nontrivial profile of the scalar and gauge fields, so-called 1RCBH \cite{Gubser:1998jb, Cvetic:1999rb, Cvetic:1999ne}. This geometry is dual to a four-	dimensional strongly coupled gauge theory N = 4 SYM at finite temperature and chemical
	potential.

\subsection{Background}
 The bulk part of the 1RCBH model is given by the following  action \cite{Gubser:1998jb}
 \begin{align}\label{EMD}
{\cal{S}}_{\textrm{bulk}}=\frac{1}{16 \pi G_5}\int d^5x\sqrt{-g}\left({\cal{R}}-\frac{f(\phi)}{4}F_{\mu\nu}F^{\mu\nu}-\frac{1}{2}(\partial_\mu\phi)^2-V(\phi)\right),
\end{align}
where $G_5$ is the $5$-dimensional Newton constant, $g$ and ${\cal{R}}$ are the determinant of the metric and its corresponding Ricci scalar, respectively, and $F_{\mu\nu}$ is the field strength of the gauge field $A_{\mu}$. The self-interacting dilaton potential $V(\phi)$ and the Maxwell-dilaton coupling $f(\phi)$ are given by
\begin{align}
&V(\phi)=-\frac{1}{L^{2}}(8e^{\frac{\phi}{\sqrt{6}}}+4 e^{-\sqrt{\frac{2}{3}\phi}}) ,\\
& f(\phi)= e^{-2\sqrt{\frac{2}{3}\phi}}.
\end{align}
where $L$ is the asymptotic $AdS_5$ radius. For simplicity, in the following we set $L = 1$. 
The equations of motion corresponding to the action \eqref{EMD} are given by 

\begin{align}\label{eq3}
 \boldsymbol{\varphi} \equiv &{1\over \sqrt{-g}} \partial_\mu\left(\sqrt{-g} g^{\mu \nu} \partial_\nu \phi\right) - {f'(\phi) \over 4} F_{\mu \nu}F^{\mu \nu} - V'(\phi)=0,\nonumber\\
 \mathcal{A}^\nu \equiv &\partial_\mu\left(\sqrt{-g} f(\phi) F^{\mu\nu}\right)=0,\nonumber\\
 \mathcal{G}_{\mu \nu} \equiv & R_{\mu \nu} - {g_{\mu \nu} \over 3} \left(V(\phi) - {f(\phi) \over 4} F_{\alpha \beta} F^{\alpha \beta}\right) - {1\over 2}\partial_\mu \phi \partial_\nu \phi - {f(\phi) \over 2} F_{\mu \alpha} F_\nu^\alpha=0.
 \end{align}

The 1RCBH solution  to the equation of motion of the action  \eqref{EMD}  is described by the following metric
\begin{align}\label{1RCBH}
ds^2=e^{2A(r)}\bigg(-h(r)dt^2+d\vec{x}^2\bigg)+\frac{e^{2B(r)}}{h(r)}dr^2
\end{align}
where
\begin{align}\label{coef}
A(r)&=\ln\left(r\right)+\frac{1}{6}\ln \left(1+\frac{Q^2}{r^2}\right),\cr 
B(r)&=-\ln\left(r\right)-\frac{1}{3}\ln \left(1+\frac{Q^2}{r^2}\right),\cr
h(r)&=1-\frac{M^2}{r^2(r^2+Q^2)}, \cr
\phi(r)&=-\sqrt{\frac{2}{3}} \ln\left(1+\frac{Q^2}{r^2}\right),\cr
a_t(r)&=\left(\frac{MQ}{r_{H}^2+ Q^2}-\frac{MQ}{r^2 +Q^2} \right),
\end{align}
where $r$ is the radial bulk coordinate, holographic direction, and the strongly coupled field theory lives on the boundary at $r \to \infty$. The time component of the gauge field is  $A_t(r)$  which is chosen to be zero on the horizon and regular on the boundary. $M$ is the black hole mass and $Q$ is its charge. $r_H$ is the black hole horizon position which is obtained from $h(r_{H}=0)$ and can be written in terms of the charge $Q$ and mass $M$ of the black hole as follow
\begin{align}
r_H=\sqrt{\frac{\sqrt{Q^4+4M^2}-Q^2}{2}}
\end{align}
The Hawking  temperature $T$ and the chemical potential  $\mu$ of the dual field theory are also given by
\begin{subequations}
\begin{align}
\label{temperature}T&=\frac{\sqrt{-g_{tt}' {g^{rr}}' }}{4\pi}\Bigg\vert_{r=r_{H}}=\left(\frac{Q^2+2r_{H}^{2}}{2\pi\sqrt{Q^2+r_H^2}}\right),\\
\label{ch}\mu &=\lim\limits_{r\rightarrow\infty}A_t=\frac{Qr_{H}}{\sqrt{Q^2+r_H^2}}.
\end{align}
\end{subequations}
where $' \equiv \frac{d}{dr}$ denotes a derivative with respect to $r$. We can now parametrize the class of solutions corresponding to the 1RCBH model by different values of the dimensionless ratio $\frac{Q}{r_H}$ which is given by  
\begin{align}
\frac{Q}{r_H}=\sqrt{2} \bigg{(} \frac{1\pm \sqrt{1-(\frac{\sqrt{2}}{\pi}\frac{\mu}{T})^2}}{\frac{\sqrt{2}}{\pi}\frac{\mu}{T}} \bigg{)},
\end{align}
implying that $0\leq\frac{\mu}{T}\leq\frac{\pi}{\sqrt{2}}$. We point out  from above equation that there are  two different values of $\frac{Q}{r_H}$ corresponding to each value of $\frac{\mu}{T}$ which parametrize two different branches of solutions. Utilizing the relations between entropy $s$ and charge density $\rho$ in terms of the bulk solution parameters $Q$ and $r_H$, we can compute the Jacobian $\mathcal{J}=\frac{\partial (s,\rho)}{\partial (T,\mu)}$. At the critical point $(\frac{\mu}{T})^*=\frac{\pi}{\sqrt{2}}$ ($\frac{Q}{r_H}=\sqrt{2}$)  the Jacobian becomes zero and  two branches intersect.  Thermodynamically stable (unstable) states correspond to the positive (negative) Jacobian branches. The branch with the parameters satisfying $\frac{Q}{r_H}<\sqrt{2}$ is stable.
 \subsection{Shock wave geometry}
 In holographic theories, the butterfly effect corresponds to the blue shift suffered by a probe particle in the bulk, which we called $W$ particle,  that falls towards the  horizon.  $W$-particle's energy is blue-shifted by the black hole's temperature for late times from the point of view of the $t=0$ slice of the geometry 
\begin{align}
E=E_{0}e^{(\frac{2\pi}{\beta})t},
\end{align}
where $E_{0}$ is the  asymptotic past energy of the $W$ particle. The back-reaction of this particle on the geometry   becomes significant at late times and  creates a shock-wave along the horizon. The effect of the shock wave is to produce a shift in the trajectory of $W$ particle. The interesting point is that the shock wave profile contains information about the parameters characterizing the chaotic behavior of the boundary theory.

We now would like to calculate  the butterfly effect velocity corresponding to the 1RCBH model \eqref{1RCBH} by constructing the relevant shock wave geometries. To do so, we assume a  general $d+1$ dimensional metric of the following form 
\begin{align}\label{general}
ds^2=-f_1(r)dt^2+f_2(r)dr^2+f_3(r)dx_i^2 , \ \ \ \ \ \ \ i=1,2,...,d-1,
\end{align}
whose boundary is located at $r\rightarrow \infty$ and  is assumed to be asymptotically $AdS_{d+1}$. We take the horizon as located at $r=r_H$ on which $f_1(r)$ vanishes and $f_2(r)$ has a first-order pole.  To describe the  shock wave solution of this geometry,  it is more useful to re-write  the above metric in the Kruskal-Szekeres coordinates. We  first define the so-called tortoise coordinate
\begin{align}
dr^*=\sqrt{\frac{f_2(r)}{f_1(r)}}dr,
\end{align}
which behaves such that $r^*\rightarrow \infty$ as $r\rightarrow r_H$, and then introduce the  Kruskal-Szekeres coordinates as follows 
\begin{align}
U=e^{\frac{2\pi}{\beta} (r^*-t)}, \ \ \ \ \  V=-e^{\frac{2\pi}{\beta} (r^*+t)},
\end{align}
  where $\beta = \frac{4 \pi}{f_1'(r_H)}$. In terms of these coordinates, metric \eqref{general}  can  be written
\begin{align}
ds^2=A(UV)dUdV+(f_3)_{ij}(UV)dx^idx^j,
\end{align}
where $ i,j=1,2,...,d-1$ run over the $d-1$ transverse directions, and $A(UV)$ is a function given by the component of the general metric as follows
\begin{align}
A(UV)=\frac{\beta^2 f_1(UV)}{4\pi ^2 UV}.
\end{align}
The horizons are also located  at $U=0$ and $V=0$, and the left and right asymptotic boundaries are positioned in $UV=-1$. These coordinates cover the maximally extended eternal black hole solution including two entangled boundaries connected by a wormhole.

As we have mentioned before, the back-reaction of $W$ particle on the geometry  becomes significant at late times, (i.e. $t>\beta$). The associated stress energy distribution becomes  highly compressed in the $V$ direction and stretched in the $U$ direction and one can approximately read the stress-energy tensor of the $W$ particle as 
\begin{align}\label{ST}
T_{VV}\sim\ \beta^{-1}\,\,\,e^{(\frac{2\pi}{\beta})t} \delta (V) a(\vec{x}),
\end{align}
where $T_{VV}$ is localized at $V=0$ and $a(\vec{x})$ is a generic function whose precise form  depends on details of the perturbation in spatial direction, as well as the propagation to the horizon. The  back-reaction of the above matter distribution is a shock wave geometry whose metric is described by
\begin{align}
ds^2=A(UV)dUdV+(f_3)_{ij}(UV)dx^idx^j-A(UV)h(t,\vec{x})\delta (V)dV^2,
\end{align}
where $h(t,\vec{x})$ is the corresponding shift in the $U$ direction,  $U\longrightarrow  U+h(t,\vec{x})$, as $W$-particle crosses the $V=0$ horizon.  Eventually, one can determine the precise form of $h(t,\vec{x})$ by solving the $V V$ component of Einstein equation
\begin{align}
R_{\mu\nu}-\frac{1}{2}g_{\mu\nu}R+\Lambda g_{\mu\nu}=8\pi G_{N} T_{\mu\nu},
\end{align}
where  $\Lambda=-\frac{d(d-1)}{2}$ is the cosmological constant, and $T_{\mu\nu}$ is given by \eqref{ST}. To obtain $h(t,\vec{x})$ one can solve the $v-v$ component of Einstein's equations by setting $\delta'(V)=\frac{-\delta(V)}{V}$ and $V^{2}\delta(V)^{2}=0$. For a local perturbation, i.e. $a(\vec{x})=\delta^{d-1}(\vec{x})$ we get
\begin{align}\label{Shift}
\left(\partial _i\partial _i-\mu ^2\right)h(\vec{x})=\frac{16\pi\, G_{N} f_3(0)}{A(0)}\beta^{-1}e^{(\frac{2\pi}{\beta})t}\,\, \delta^{d-1}(\vec{x}).
\end{align}
Assuming $(f_3)_{ij}$ to be diagonal and isotropic, the solution to the above differential equation  at long distances $x\gg \mu^{-1}$ reads
\begin{align}\label{shock}
h(x)\sim \frac{e^{(\frac{2\pi}{\beta})(t-t_{\ast})-\mu \vert x \vert}}{\vert x \vert^{\frac{d-1}{2}}},
\end{align}
where the screening length $\mu $ is  given by
\begin{align}
\label{screening}\mu ^2=\frac{(d-1)}{A(0)}\frac{\partial f_3(UV)}{\partial (UV)}\bigg\vert_{V=0},
\end{align}
and the scrambling time $t_{\ast} \sim \frac{\beta}{2\pi} \log \frac{1}{G_{N}}$. For later convenience, we express $\mu$ as a function of  the coordinates $ (t, r)$ 
\begin{align}\label{mu}
\mu ^2=\frac{(d-1)\pi f'_3(r)}{\beta \sqrt{f_1(r) f_2(r)}}\bigg\vert_{r=r_H},
\end{align}
where  $' \equiv \frac{d}{dr}$ is the radial derivative. For the case of the $d+1-$ dimensional AdS-Schwarzchild  in which $f_1(r)=r^2-r^{2-d}$, $f_2(r)=f_1(r)^{-1}$, $r_H=1$ and $\beta=\frac{4\pi}{d}$ one finds $\mu ^2=\frac{d(d-1)}{2}$ as expected \cite{Roberts:2014isa}. On the other hand, the  shock wave profile \eqref{shock} contains information regarding the parameters such as  $\lambda_{L}$ and   $v_B$ which are corresponded to  the chaotic behavior of the boundary theory. Hence, one can  read off  $\lambda_{L}$ and $v_B$ holographically as
\begin{align}\label{L}
\lambda_{L}=\frac{2\pi}{\beta},\,\,\,\,\,\,\,\,\,\,v_B=\frac{2\pi}{\beta \mu},
\end{align}
where  $\lambda_{L}$ is universal in all such theories, while $v_B$ is model dependent. Now we are in a position to extract butterfly velocity for metric \eqref{general}. Using \eqref{L} and \eqref{mu} we obtain $v_B$  as the following form
\begin{align}\label{bv}
v_B=\frac{2\sqrt{\pi}(f_1f_2)^{\frac{1}{4}}}{\sqrt{\beta (d-1)f_3'}}\bigg\vert_{r=r_H},
\end{align}
 where we simply reach the $d+1-$ dimensional AdS-Schwarzchild  result i.e., $v_B=\sqrt{\frac{d}{2(d-1)}}$, as reported in \cite{Roberts:2014isa}.

\subsection{$v_B$ in 1RCBH model}
\begin{figure}
\centering
\includegraphics[scale=0.5]{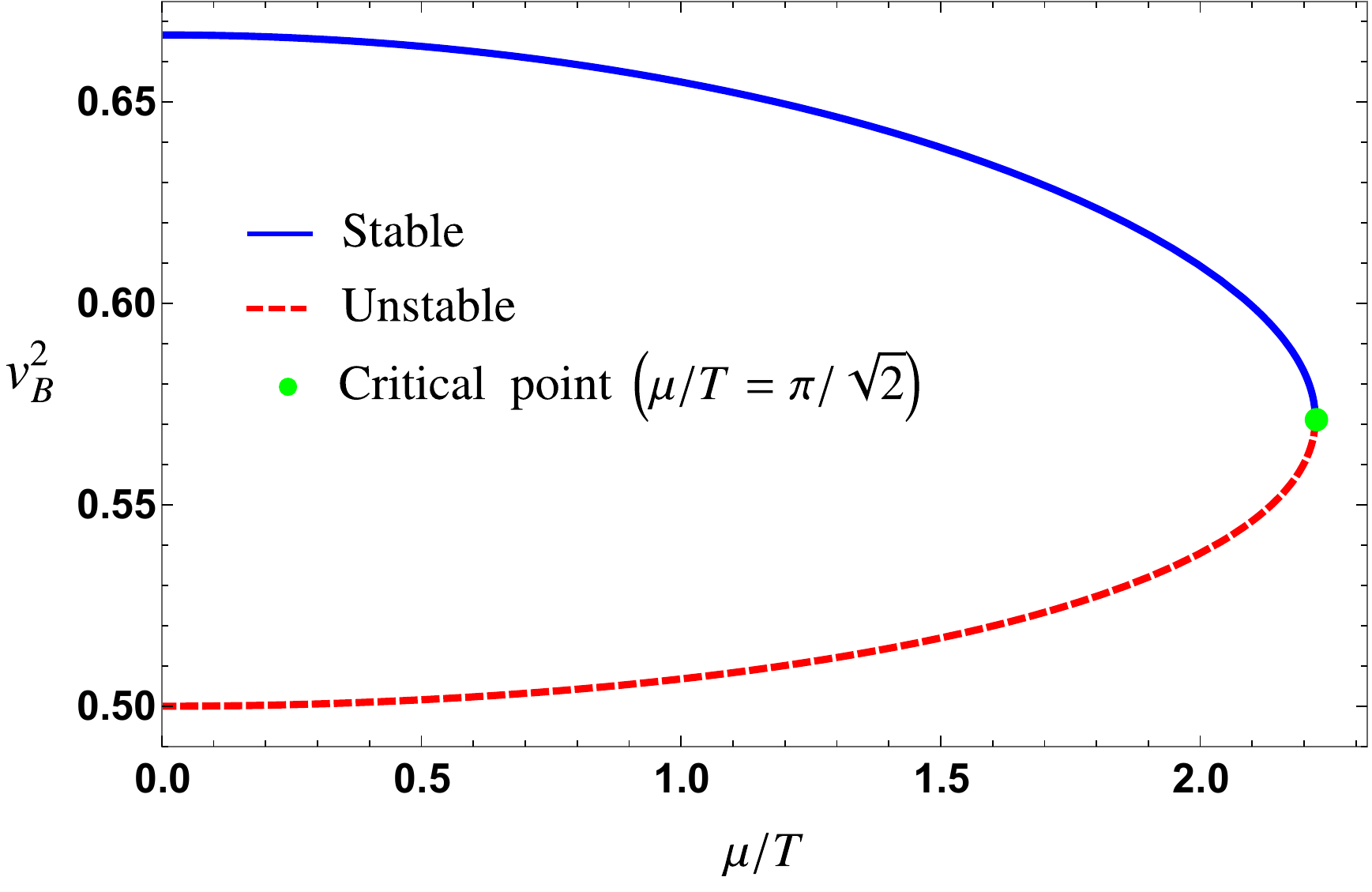}
\caption{The square of the butterfly velocity $v_B^2$ as a function of dimensionless ratio $\frac{\mu}{T}$. The solid blue curve corresponds to the stable solutions while the red dashed curve corresponds to the unstable solutions. The green circle point where these branches of solutions merge is called the critical point where $(\frac{\mu}{T})^*=\frac{\pi}{\sqrt{2}}$. }
\label{VB}
\end{figure}

Having discussed the characteristic parameters of chaotic systems such as Lyapunov exponent $\lambda_{L}$ and butterfly velocity $v_B$ in detail and obtained the explicit expressions for them, we are now in a position to write $v_B$ for 1RCBH model whose metric is given by \eqref{1RCBH}. Using \eqref{temperature}, \eqref{ch}, \eqref{coef} and \eqref{bv} and considering $f_3(r)=e^{2A(r)}$, $f_1(r)=h(r) f_3(r)$ and $f_2(r)=\frac{e^{2B(r)}}{h(r)}$, one can read the butterfly velocity of 1RCBH model  as follows
\begin{align}
v_B^2=\frac{4}{7\mp \sqrt{1-\left(\frac{\mu /T}{\pi /\sqrt{2}}\right)^2}},
\end{align}
where $-(+)$ indicates the stable(unstable) black hole solutions. The above relation has been depicted in Fig. \ref{VB} for both stable and unstable black hole solutions for different values of $\frac{\mu}{T}$. The circle point has specified the precise location of the critical point which is at $(\frac{\mu}{T})^*=\frac{\pi}{\sqrt{2}}$. The point is that for all values of $\frac{\mu}{T}$, we observe that $v_B^2 > c_s^2 = \frac{1}{3}$ where $c_s$ is the speed of sound wave propagations \footnote{This is the result we obtained for two models in which the conformal symmetry is respected and hence this is not generic in some sense. One can find some models which are not conformal and this bound is violated, see for example \cite{Baggioli:2020ljz}.}. Also, $v_B$ gets smaller (larger) towards the critical point in stable (unstable) branches solution signaling the fact that the speed of information propagation can be a good diagnostic quantity to pinpoint the location of critical point. 

\section{Critical exponent}\label{sec3}
As mentioned before, the 1RCBH model we study in this paper enjoys a critical point at $(\frac{\mu}{T})^*=\frac{\pi}{\sqrt{2}}$. The behavior of different observables near the critical point has been studied in the literature \cite{DeWolfe:2011ts,Ebrahim:2018uky,Ebrahim:2020qif,Amrahi:2020jqg,Amrahi:2021lgh,Finazzo:2016psx,Ebrahim:2017gvk,Asadi:2021hds} in which the authors  have shown that this behavior can be considered as $\left((\frac{\mu}{T})^*-\frac{\mu}{T}\right)^{-\theta}$ where $\theta$ is the dynamical exponent which is obtained to be $\frac{1}{2}$. We also would like to  study the behavior of butterfly velocity $v_B$ near this critical point. To proceed, we expand $v_B$ and  $\frac{dv_B}{d\left(\frac{\mu}{T}\right)}$ in power of $\left((\frac{\mu}{T})^*-\frac{\mu}{T}\right)$ as follows
 \begin{align}
v_B=\frac{2}{\sqrt{7}}+\frac{2^{3/4}}{7\sqrt{7\pi}}\left((\frac{\mu}{T})^*-\frac{\mu}{T}\right)^{\frac{1}{2}}+\mathcal{O}\left[\left((\frac{\mu}{T})^*-\frac{\mu}{T}\right)^1\right],
 \end{align}
 \begin{align}
\frac{dv_B}{d\left(\frac{\mu}{T}\right)}=\frac{1}{7\times 2^\frac{1}{4}\sqrt{7\pi}}\left((\frac{\mu}{T})^*-\frac{\mu}{T}\right)^{-\frac{1}{2}}+\mathcal{O}\left[\left((\frac{\mu}{T})^*-\frac{\mu}{T}\right)^0\right].
 \end{align}
It has been  seen that  $v_B$ remains finite, while $\frac{dv_B}{d\left(\frac{\mu}{T}\right)}$ diverges near the critical point and  one can see that the dynamical exponent is equal to $\frac{1}{2}$. This also shows that butterfly velocity can probe the location of critical point and in another word can be used as a prob to investigate the phase diagram of the 1RCBH model. The reason why we look for  $v_B$ to probe the critical point is because this is a dynamical quantity which has a time scale and seems to have such scaling property.\footnote{We are grateful to Saso Grozdanov for his valuable comment regarding this subject.}

\section{Pole-skipping in 1RCBH model}\label{sec4}
As we have pointed out before,  OTOCs can be recognized as very useful tools to diagnose many-body quantum chaos. In addition to OTOCs, the chaotic nature of many-body thermal systems is also encoded in the equations of motion for linearized fluctuations in the spin-0 sector. For a general and non-critical background, it has been shown that the $vv$ component of Einstein's equation near the horizon gets the following form \cite{Blake:2018leo}
\begin{align}
	\left(k^2 + k_0^2\frac{\omega}{2 \pi T i}\right) \delta g^{0}_{v v}(r_H) + (\omega - 2 i \pi T) \left(\omega \delta g^{0}_{x^i x^i}(r_H) + 2 k \delta g^{0}_{v z}(r_H)\right) = 0,
\end{align}
if we parametrize the propagations along the $z$ direction. At chaos point, i.e. $\omega = 2 \pi T i$ and $k = i k_0$ this equation becomes identically zero. A consequence of this issue is that there exist plenty of linearly independent solutions to the Einstein equations. All these linear solutions approach the chaos point by different slope 
	\begin{align}
		\frac{\delta \omega}{\delta k} = \frac{2 k_0 \delta g^{0}_{v v}(r_H)}{\frac{k_0^2}{\omega_0} \delta g^{0}_{v v}(r_H) - \omega_0 \delta g^{0}_{x^i x^i}(r_H) - 2 k_0 \delta g^{0}_{v z}(r_H)},
	\end{align}
	where $\omega_0 = 2 \pi T$. Another manifestation  of having many solutions is that two regular solutions exist in ingoing coordinates, i.e. solutions become $(r -r_H)^{\alpha_1, \alpha_2}$ where both $ (\alpha_1, \alpha_2)\geq 0$. In this section, we want to examine these features for the 1RCBH model.


\subsection{Setup for 1RCBH model}
In this subsection, we would like to study the linearized equations of motion on top of the metric  \eqref{general} near the following  points
\begin{align}
\lambda_{L}=\frac{2\pi}{\beta}=2\pi T, \,\,\,\,\,\,\,\,\,\,\,\,\,(k_{0})^{2}=\frac{\pi (d-1)f_3'(r_H)}{\beta \sqrt{f_1(r_H)f_2(r_H)}},
\end{align}
where $k_{0}$ has been calculated through \eqref{bv} and $k_{0}=\frac{\lambda_{L}}{v_B}$.  
The sound mode which corresponds to the retarded hydrodynamic correlators obeys the equations whose solutions satisfy the ingoing wave boundary conditions at the horizon. Note also that  these solutions in part contain the $vv$ component of Einstein’s equations  where $v$ is the ingoing Eddington-Finkelstein (EF) coordinate. It is therefore convenient to introduce ingoing  EF coordinates ($v; r; x^{i}$) in terms of which the general metric \eqref{general} becomes 
\begin{align}\label{ef}
ds^2=-f_1(r)dv^2+2\sqrt{f_1(r)f_2(r)}dvdr+f_3(r)(dx^{i})^2.
\end{align}
 Under infinitesimal diffeomorphism $x^{\mu}\rightarrow x^{\mu}+\xi^{\mu}$ and an infinitesimal gauge transformation $A_{\mu}\rightarrow A_{\mu}+ \nabla_{\mu}\Lambda$, the perturbations $\delta g_{\mu \nu}$, $\delta A_{\mu }$ and $\delta \phi$ transform  as
\begin{align}
&\delta g_{\mu \nu}\rightarrow \delta g_{\mu \nu}- \nabla_{\mu}\xi _{\nu}-\nabla_{\nu}\xi _{\mu}, \no\\ 
&\delta A_{\mu}\rightarrow \delta A_{\mu}+ \nabla_{\mu} \Lambda - \xi ^{\nu}\nabla_{\nu} A_{\mu}- A_{\nu}\nabla_{\mu}\xi^{\nu}, \no\\
&\delta \phi \rightarrow \delta \phi -\xi ^{\nu} \nabla_{\nu} \phi , \no
\end{align}
where $\xi _{\mu}$ and $\Lambda$ are diff and gauge functions, respectively. 
Due to the symmetry considerations, we  set the   metric, gauge and scalar field perturbation along the $x^{3}\equiv z$ direction as follows
\begin{align}\label{setfluc}
	&\delta g_{\mu \nu}(v, r, z) = e^{- i \omega v + i k z} \delta g_{\mu \nu}(r), \no\\
	&\delta A_{\mu }(v, r, z) = e^{- i \omega v + i k z} \delta A_{\mu }(r),\no\\
	&\delta \phi(v, r, z) = e^{- i \omega v + i k z} \delta \phi(r),
\end{align}
Above perturbations can be classified  into three sets of sectors named scalar (spin 0), vector (spin 1) and tensor (spin 2) sectors respectively, which are decoupled at the level of equations due to the symmetry \cite{Jansen:2019wag}. Fluctuations in the spin 0 sector and EF coordinates are given by
\begin{align}
\delta \Phi = \bigg(\delta g_{v v}, \delta g_{r v}, \delta g_{rr}, \delta g_{r z}, \delta g_{v z}, \delta g_{x^{i}x^{i}}, \delta A_{v}, \delta A_{z}, \delta \phi \bigg).
\end{align}
which due to the $SO(3)$ invariance,  $\delta g_{y y} = \delta g_{x x}$. The linearized equations for these perturbations  derived from equations \eqref{eq3} are very hard to solve. But  series solution exist for every single point of the bulk.  According to the fluid/gravity conjecture, the small energy (charge) perturbations on top of the living strongly coupled theory on the boundary  correspond to equivalent small metric (gauge) field perturbations on the event horizon surface. Therefore, we scrutinize the near horizon points. We  make  series assumptions for each perturbating field as
	\begin{align}\label{expansion}
	&\delta g_{\mu \nu}(r) = \sum_{n = 0}^{\infty} \delta g_{\mu \nu}^{n}(r_H)\, (r -r_H)^n,\no\\
	&\delta A_{\mu}(r) = \sum_{n = 0}^{\infty} \delta A_{\mu}^{n}(r_H)\, (r -r_H)^n,\no\\
	&\delta \phi(r) = \sum_{n = 0}^{\infty} \delta\phi^{n}(r_H)\, (r -r_H)^n,
	\end{align}
	and plug them into the linearized equations to solve the  series order by order in $\epsilon = r -r_H$.  We observe that  the $\mathcal{G}_{v v}$ equation for fluctuations near the horizon on top of the 1RCBH background gets the following form
	\begin{align}\label{eqgvv}
	\delta g_{v v}^{0}(r_H) \left( k^2 - \frac{i k_{0}^2 \omega}{2 \pi T} \right)+ (\omega -2 i \pi  T) (2 q \delta g_{z v}^{0}(r_H)+2 \omega  \delta g_{x x}^{0}(r_H)+\omega  \delta g_{z z}^{0}(r_H)) = 0,
	\end{align} 
	where $k_{0}^{2} = \frac{\lambda_{L}^2}{v_B^2} = 6 \pi T A'(r_H) e^{A(r_H)-B(r_H)}$. For general $\omega$ and $k$  equation \eqref{eqgvv} imposes a non-trivial constraint on the  near-horizon expansion  components $\delta g_{v v}^{0}(r_H)$, $\delta g_{z v}^{0}(r_H)$, $\delta g_{x x}^{0}(r_H)$ and $\delta g_{z z}^{0}(r_H)$. But  at the point  $\omega_{*}=i\lambda_{L}=2 \pi Ti$ the metric component $\delta g_{v v}^{0}(r_H)$ decouples from the other components and equation \eqref{eqgvv}  reduces to
\begin{align}\label{eqgvvred}
	\delta g_{v v}^{0}(r_H) \left( k^2 - \frac{i k_{0}^2 \omega}{2 \pi T} \right) = 0.
\end{align} 
Interestingly,  this equation has the same form as the equation that determines the profile of shift corresponding to  the  shock wave geometry, equation \eqref{Shift}, with  $\mu^{2}=\frac{\lambda_{L}^2 }{v_B^2}$. The other point is that at the point $k=i k_{0}$  equation \eqref{eqgvvred}  is automatically satisfied. Put in other words,  at chaos point, $\omega = 2 i \pi T$ and $k = i k_0$ equation \eqref{eqgvv} becomes trivial and it does not impose any constraint on the near-horizon expansion components  $\delta g^{0}_{v v}(r_H)$, $\delta g^{0}_{z v}(r_H)$, $\delta g^{0}_{x x}(r_H)$ and $\delta g^{0}_{z z}(r_H)$. Its message is that there exists one extra ingoing mode emerging at this point and as a result lead to the pole-skipping in  the retarded energy density correlation function $G^{R}_{T^{00}T^{00}}(\omega,k)$  at the chaos point. To understand this phenomenon more precisely, we  can now check what happens as one moves slightly away from the special point. We consider Einstein’s equations with $\omega=i\lambda+\epsilon\,\delta\omega$ and $k=i k_{0} + \epsilon\, \delta k$ where $\vert \epsilon \vert\ll 1$ and hence, at leading order in $\epsilon$,  we have a  family  of different ingoing modes by varying the slope $\frac{\delta \omega}{\delta k}$. This slope can be chosen to correspond to the resulting ingoing mode near the chaos point with different asymptotic solutions at the boundary. If one chooses $\frac{\delta \omega}{\delta k}$ in the following form 
	\begin{align}\label{eqslope}
	\frac{\delta \omega}{\delta k} = \frac{2 k_{0} \delta g^{0}_{v v}(r_H)}{\frac{k_{0}^2}{2 \pi T}\delta g^{0}_{v v}(r_H) - 2 k_{0} \delta g^{0}_{z v}(r_H) - 2 \pi  T \left(2\delta g^{0}_{x x}(r_H) + \delta g^{0}_{z z}(r_H)\right)},
	\end{align}
then we will get an ingoing mode that matches continuously onto the normalizable solution at the boundary. All these lines pass through the chaos point and if we move away from that point, we see different lines with slopes \eqref{eqslope} in the $T_{v v}$ correlation functions. This can also be described as a line of poles in the energy density correlation function that passes through the special point when we move away from the special point along the slope \eqref{eqslope}. The point is that one could also choose  different $\frac{\delta \omega}{\delta k}$ such that the ingoing mode matches onto a solution with different asymptotes at the boundary. 

Apart from this viewpoint, we can study linearized equations  more fashionably. To do this, we have to decouple the linearized equations by choosing the suitable gauge+diff invariant master field variables. In the spin 0 sector, there are 7 combinations of gauge+diff-invariant perturbations \cite{Jansen:2019wag,Kodama:2003jz}, but in the on-shell level only three of them are independent. In the following, we write a specific choice of these three combinations for the general metric
\begin{align}\label{eq47}
	\psi_1 &\equiv \frac{d}{dr}\left[\frac{2\delta g_{x x}(r)  + \delta g_{z z}(r)}{f_3(r)}\right] - \frac{i \omega}{f_3(r)} \sqrt{\frac{f_2(r)}{f_1(r)}} \left(2\delta g_{x x}(r)  + \delta g_{z z}(r)\right)\no\\
	& - \frac{2 i k}{f_3(r)} \left(\delta g_{r z}(r) + \sqrt{\frac{f_2(r)}{f_1(r)}}\delta g_{v z}(r)\right) - \frac{3 f_3'(r)}{2 f_3(r) f_2(r)} \left(\delta g_{rr}(r) + 2 \sqrt{\frac{f_2(r)}{f_1(r)}}\delta g_{r v}(r) + \frac{f_2(r)}{f_1(r)} \delta g_{v v}(r)\right) \no\\
	& - \left(2 k^2\frac{f_2(r)}{f_3(r) f_3'(r)} + \frac{3}{ f_3(r)} \left(\frac{f_3''(r)}{f_3'(r)}  -  \frac{f_3'(r)}{f_3(r)} - \frac{f_2'(r)}{2 f_2(r)}\right)\right) \delta g_{x x}(r) ,\no\\
	\psi_2 & \equiv \frac{\delta g_{x x}(r)}{f_3'(r)} \phi'(r) - \delta \phi(r),\no\\
	\psi_3 &\equiv k \delta A_v(r) + \omega \delta A_z(r) - \frac{k A_t'(r)}{f_3'(r)} \delta g_{x x}(r).
\end{align}
In appendix \ref{Appendix1} we discuss the independent combinations of gauge+diff-invariant perturbations in detail. By choosing $(f_1(r), f_2(r), f_3(r)) = (e^{2 A(r)} h(r), \frac{e^{2 B(r)}}{h(r)}, e^{2 A(r)})$ in the equation \eqref{ef}, we rewrite  the 1RCBH metric  in the  EF coordinates  
\begin{align}\label{ef1RCBH}
ds^2=-e^{2A(r)}h(r)dv^2+2e^{A(r)+B(r)}dvdr+e^{2A(r)}dx_i^2.
\end{align}
So, we will get the following gauge-invariant master field variables
\begin{align}\label{gi1RCBH}
	\psi_1 &\equiv \frac{d}{dr}\left[e^{- 2 A(r)}\left(2\delta g_{x x}(r)+ \delta g_{z z}(r)\right)\right] - \frac{i \omega e^{B(r) - 3 A(r)}}{h(r)}  \left(2\delta g_{x x}(r) + \delta g_{z z}(r)\right)\no\\
	& - 2 i k e^{- 2 A(r)} \left(\delta g_{r z}(r) + \frac{e^{B(r) -  A(r)}}{h(r)}\delta g_{v z}(r)\right) \no\\
	&- 3 h(r) A'(r) e^{- 2 B(r)}\left(\delta g_{rr}(r) + 2 \frac{e^{B(r) -  A(r)}}{h(r)}\delta g_{r v}(r) + \frac{e^{2B(r) -  2 A(r)}}{h^2(r)} \delta g_{v v}(r)\right) \no\\
	& - \left( k^2\frac{e^{2 B(r) - 4 A(r)}}{A'(r) h(r)} + \frac{e^{-2 A(r)} \left( 3 A'(r) h'(r) + h(r) (6 A''(r) - 6 A'(r) B'(r))\right)}{ 2 A'(r) h(r)} \right) \delta g_{x x}(r),\no\\
	\psi_2 & \equiv e^{- 2 A(r)}\frac{\delta g_{x x}(r)}{2 A'(r)} \phi'(r) - \delta \phi(r),\no\\
	\psi_3 &\equiv k \delta A_v(r) + \omega \delta A_z(r) - k A_t'(r) e^{- 2 A(r)}  \frac{\delta g_{x x}(r)}{ 2 A'(r)}.
\end{align}
 The linearized perturbation equations can  be written  for these gauge+diff-invariant perturbations.  But  the choices of metric perturbations in master field variables are not correct, since they are rank (0, 2) tensors and vary under coordinate transformations. We have to convert them to the rank (1, 1) tensors and then plug them into these master equations. To do so, we note that $\delta g_{\mu \nu} = g^0_{\mu \alpha} \delta g^\alpha_\nu$ where $g^0_{\mu \alpha}$ are metric components written in \eqref{ef1RCBH}. Hence, we have
\begin{align}
	&\delta g_{r r} = e^{A(r)+B(r)} \delta g^v_r, \qquad \delta g_{r v} =  e^{A(r)+B(r)} \delta g^v_v = - e^{2 A(r)} h(r) \delta g^v_r +  e^{A(r)+B(r)} \delta g^r_r, \nonumber\\
	& \delta g_{v v} = - e^{2 A(r)} h(r) \delta g^v_v +  e^{A(r)+B(r)} \delta g^r_v, \qquad \delta g_{i i} = e^{2 A(r)} \delta g^i_i, \,\,\, i = x, z, \nonumber\\
	& \delta g_{z j} = e^{2 A(r)} \delta g^z_j, \,\,\, j = r, v.
\end{align}
These replacements should be done in the invariant combinations of equations \eqref{gi1RCBH}. To solve the linearized equations, it is useful to consider  the two points. First, from the invariant sets of the equations \eqref{gi1RCBH} or any combinations, we omit $(\delta g^r_r, \delta A_v, \delta \phi)$ in favor of other fluctuations. Second, the zero-order solutions are used to relate higher derivatives of functions $(A^{(n)}(r), h^{(n)}(r), \phi^{(n)}(r), \cdots)$ with $n \geq 2$ to the lower orders $(A^{(n)}(r), h^{(n)}(r), \phi^{(n)}(r), \cdots)$ with $n \leq 1$. The Number of independent equations is 11. Four of them are constrained equations that only contain first-order derivative to the $"r"$ coordinate and the rests are dynamical equations with two and one order of derivatives. To obtain the constrained equations, we define a normal vector $n_\mu = (0, 1, 0, 0, 0)$ to the $r= cte$ hypersurfaces. 
 Hence, the constrained equations are
\begin{align}
	&n^\mu \mathcal{G}_{\mu \nu} = g^{\mu \alpha} n_\alpha \mathcal{G}_{\mu \nu} = 0, \nonumber\\
	& n^\mu \mathcal{A}_{\mu} = g^{\mu \alpha} n_\alpha \mathcal{A}_{\mu} = 0.
\end{align}   
For the 1RCBH background, these equations can be written as follows
\begin{align}\label{eq1d}
	&n^v \mathcal{G}_{v v}  + n^r \mathcal{G}_{r v} =  e^{- A(r) - B(r)} \mathcal{G}_{v v} + e^{- 2 B(r)} h(r) \mathcal{G}_{r v}= 0, \nonumber\\
	&n^v \mathcal{G}_{r v}  + n^r \mathcal{G}_{r r} =  e^{- A(r) - B(r)} \mathcal{G}_{v r} + e^{- 2 B(r)} h(r) \mathcal{G}_{r r}= 0, \nonumber\\
	& n^v \mathcal{G}_{v z}  + n^r \mathcal{G}_{r z} =  e^{- A(r) - B(r)} \mathcal{G}_{v z} + e^{- 2 B(r)} h(r) \mathcal{G}_{r z}= 0, \nonumber\\
	&n^v \mathcal{A}_{v}  + n^r \mathcal{A}_{r} =  e^{- A(r) - B(r)} \mathcal{A}_{v} + e^{- 2 B(r)} h(r) \mathcal{A}_{r}= 0.
\end{align}
From these equations, the fluctuations $\left(\delta g'^{z}_v(r), \delta g'^{x}_x(r), \delta g'^z_z(r), \delta A'_z(r)\right)$ are solved in terms of the set 
$\left(\delta g^v_r(r), \delta g^r_v(r), \delta g^z_r(r), \delta g^z_v(r), \delta g^x_x(r), \delta g^z_z(r) \right)$ and invariant perturbations\\
 $\left(\psi_1(r), \psi_2(r), \psi_3(r), \psi'_1(r), \psi'_2(r), \psi'_3(r)\right)$. On the other hand, the dynamical equations are given by
\begin{align}
	&\boldsymbol{\varphi} = 0, \nonumber\\
	&\mathcal{A}_{r} = 0, \qquad \mathcal{A}_{z} = 0\nonumber\\
	& \mathcal{G}_{r v} = 0, \qquad \mathcal{G}_{x x} = 0, \qquad \mathcal{G}_{z z} = 0, \qquad \mathcal{G}_{v z} = 0.
\end{align}
To rewrite these dynamical equations in terms of the invariant perturbations, we have to do the following steps:
\begin{itemize}
	\item[1)] Solving the  equations $\left( \mathcal{G}_{x x} = 0, \,\,\, \mathcal{G}_{z z} = 0, \,\,\, \mathcal{G}_{v z} = 0, \,\,\, \mathcal{A}_{z} = 0\right)$ would give us the results  for $(\delta g''^x_x(r), \delta g''^z_z(r), \delta g''^z_v(r), \delta A''_z(r))$ in terms of other perturbations. Remember that from equations \eqref{eq1d}, the perturbations  $\left(\delta g'^{z}_v(r), \delta g'^{x}_x(r), \delta g'^z_z(r), \delta A'_z(r)\right)$ are solved in terms of other non-derivatives perturbations. So the set $(\delta g''^x_x(r), \delta g''^z_z(r), \delta g''^z_v(r), \delta A''_z(r))$ are written solely in terms of non-derivatives fluctuations.
	\item[2)] The solutions for $(\delta g'''^x_x(r), \delta g'''^z_z(r))$ are derived from derivatives of $"r"$ of the solutions in item one. These  are needed for simplifications of the equation $\mathcal{G}_{r v} = 0$.
	\item[3)] We insert the solutions of $(\delta g''^x_x(r), \delta g''^z_z(r), \delta g''^z_v(r), \delta A''_z(r))$ and \\
	$\left(\delta g'^{z}_v(r), \delta g'^{x}_x(r), \delta g'^z_z(r), \delta A'_z(r)\right)$ into the following equations
	\begin{align}
		&\boldsymbol{\varphi} = 0, \nonumber\\
		&\mathcal{A}_{r} = 0,\nonumber\\
		& \mathcal{G}_{r v} = 0, 
	\end{align}
	as well as using  zero-order equations. Careful simplifications would give us three dynamical equations for $\left(\psi_1, \psi_2, \psi_3\right)$.
\end{itemize}
  Symbolically, the resulting equations are as follows
	\begin{align}\label{eqinvariant}
		&f^1_1(r) \psi_1(r) + f^2_1(r) \psi_2(r) + f^3_1(r) \psi_3(r) +f^1_2(r) \psi_1'(r) +f^2_2(r) \psi_2'(r) +f^3_2(r) \psi_3'(r) +f^2_3(r) \psi_2''(r) = 0,\nonumber\\
		&g^1_1(r) \psi_1(r) + g^2_1(r) \psi_2(r) + g^3_1(r) \psi_3(r) +g^1_2(r) \psi_1'(r) +g^2_2(r) \psi_2'(r) +g^3_2(r) \psi_3'(r) +g^3_3(r) \psi_3''(r) = 0,\nonumber\\
		&m^1_1(r) \psi_1(r) + m^2_1(r) \psi_2(r) + m^3_1(r) \psi_3(r) +m^1_2(r) \psi_1'(r) +m^2_2(r) \psi_2'(r) +m^3_2(r) \psi_3'(r) \nonumber\\
		&m^1_3(r) \psi_1''(r) + m^2_3(r) \psi_2''(r) + m^3_3(r) \psi_3''(r) = 0.
	\end{align}
 where $(f^i_j, g^i_j, m^i_j)$ are some coefficients. \footnote{Since the coefficients $(f^i_j, g^i_j, m^i_j)$ are very complicated and lengthy, we pack them in three MATHEMATICA files and are provided along with this paper.}   This  procedure can be applied to any configuration of perturbating fields. 
 
 The pole-skipping feature can be examined by near-horizon behavior  of the equations \eqref{eqinvariant}. Let us call the solutions at generic $\omega$ and $k$ near the horizon as follows
\begin{align}\label{eq419}
	\Psi = a_1(\omega, k)\epsilon^{\alpha_1} + a_2(\omega, k)\epsilon^{\alpha_2}.
\end{align}
Our calculations show that the exponents are $(\alpha_1, \alpha_2) = (\frac{i \omega}{2 \pi  T}, 0)$ at generic  $\omega$ and $k$ for all perturbation fields. The solutions with $\alpha_1 = \frac{i \omega}{2 \pi  T}(\alpha_2 = 0)$ corresponds to the singular (regular) outgoing (ingoing) mode. However, at the chaos point  the story would change. The exponents are shifted by one, that is $(\alpha_1, \alpha_2) = (1+\frac{i \omega}{2 \pi  T}, 1)$,  and therefore represent two regular ingoing solutions. Unfortunately, we don't catch this property in three combinations of equations \eqref{eqinvariant}, but the point is that the appearance of equations \eqref{eqgvv} guarantees the existence of a combination of invariant perturbations which do this job.

The multivaluedness of boundary retarded Green's function can also be seen from another point of view. Recently, there have been reported that  at higher Matsubara frequencies, i.e. $\omega = \omega_n = - 2 i n \pi T$  the equations of motion of  perturbations exhibit a pole-skipping in  the $G^R_{\mathcal{O}  \mathcal{O}}(\omega, k)$ for the corresponding boundary operator\cite{Grozdanov:2019uhi, Grozdanov:2023txs, Blake:2019otz}. This is because at these points the equations give no constraints on $\delta \phi^n(r_H)$ coefficients of the expansion \eqref{expansion} and these many unknown coefficients reflect many hydrodynamics poles which each of them approaches  to that point with a special slope \cite{Blake:2019otz}. Another issue is that at $k = i k_0$ and $\omega = \omega_n$ the regular solutions are labelled with two independent parameters, since $\mbox{det}\, \mathcal{M}^n(\omega_n, k_0) = 0$ and there exist no non-trivial solution  there. We would like to explore these features in our model.

To do this, we expand the equations of motion for scalar perturbation around the horizon. The result is as follows
\begin{align}\label{pole}
	\mathcal{I}_{1} & = M_{1 1}(\omega, k^2) \delta \phi^0(r_H) + (2 \pi T - i \omega) \delta \phi^1(r_H),\nonumber\\
	\mathcal{I}_{2} & = M_{2 1}(\omega, k^2) \delta \phi^0(r_H) + M_{2 2}(\omega, k^2) \delta \phi^1(r_H) + (4 \pi T - i \omega) \delta \phi^2(r_H),\\
	\mathcal{I}_{3} & = M_{3 1}(\omega, k^2) \delta \phi^0(r_H) + M_{3 2}(\omega, k^2) \delta \phi^1(r_H) + M_{3 3}(\omega, k^2) \delta \phi^2(r_H) +  (6 \pi T - i \omega) \delta \phi^3(r_H),\nonumber
\end{align}
where the coefficients $M_{i j}(\omega, k^2)$ take the following form
\begin{align}
	M_{i j}(\omega, k^2) = i \omega a_{i j} + k^2 b_{i j} + c_{i j},
\end{align}
with $a_{i j}, b_{i j}, c_{i j}$ are determined by the background solutions in \eqref{coef} and their derivatives on the horizon. Their special form is very complicated and we list them in appendix \ref{Appendix3}. Expressions $\mathcal{I}_{1}, \mathcal{I}_{2}, \mathcal{I}_{3}$ are combinations of other perturbations with specific coefficients derived from  background solutions. They are so lengthy and not valuable for our demands.

We observe  from equations \eqref{pole} that at frequencies $\omega = \omega_n$ it is not possible to read the coefficients iteratively from $\delta \phi^0(r_H)$. It means that $\delta \phi^n(r_H)$ are no longer dependent parameters and thus are free parameters near the horizon. Also, at the point $\omega = \omega_n$ the first $n$ equation are decoupled and we can solve a simple matrix equation for $\delta \tilde{\phi} = \left(\delta \phi^0(r_H), \cdots, \delta \phi^{n-1}(r_H)\right)$ as it follows
\begin{align}
	\mathcal{M}^{n}(\omega, k^2) \cdot \delta \tilde{\phi}= \mathcal{I}.
\end{align} 
This feature is similar to former observation \cite{Grozdanov:2019uhi, Grozdanov:2023txs, Blake:2019otz}. However, we observe that at $k = i k_0$ and $\omega = \omega_n$, $\mbox{det}\, \mathcal{M}^n(\omega_n, k_0) \neq 0$. Therefore, solutions for the linear equations \eqref{pole} are labelled with just one parameter. This is the novelty of our computations which may be because of the presence of the electric charge. It is worthwhile to mention that equations for gauge field perturbations do not capture such properties.
\subsection{ Pole skipping in AdS-RN(AdS-Reisner-Nordstrom) black hole}
As an extra example, we provide linearized equations of perturbation on top of the AdS-RN background that is a charged black hole and according to the AdS/CFT correspondence is dual to a strongly coupled field theory with a $U(1)$ charge. The bulk action of AdS-RN action is given by
\begin{align}
	S_{\textrm{AdS-RN}} = \frac{1}{16 \pi G_5} \int\, d^5x\, \sqrt{-g} \bigg(\mathcal{R} - \frac{1}{4} F_{\mu \nu} F^{\mu \nu}\bigg).
\end{align}
Equations of motion for the metric and gauge fields are as follows
\begin{align}
	&\nabla_{\mu} F^{\mu \nu} = \frac{1}{\sqrt{-g}} \partial_\mu \left(\sqrt{-g} F^{\mu \nu}\right)= 0, \no\\
	& \mathcal{R}_{\mu \nu} +  g_{\mu \nu} \bigg(4 + \frac{1}{12} F_{\alpha \beta} F^{\alpha \beta}\bigg) - \frac{1}{2} F_{\mu \alpha} F_{\nu}^\alpha = 0.
\end{align}
Solutions to these equations of motion are 
\begin{align}\label{soladsrn}
	&ds^2 = - h(r) dt^2 + \frac{1}{h(r)} dr^2 + r^2 d\vec{x}^2,\no\\
	&A_\mu = a_t(r) dt,
\end{align}
where
\begin{align}
	&h(r) = r^2 \left(1 - \frac{r_H^4}{r^4}\right) + M \left(\frac{r_H^2}{r^4} - \frac{1}{r^2}\right),\no\\
	&a_t(r) =  Q\left(1 - \frac{r_H^2}{r^2} \right),
\end{align}
with $M = \frac{Q^2 r_H^2}{3}$. The Hawking temperature  and the $U(1)$ charge of the dual field are also given by
\begin{align}
	&T = \frac{f'(r_H)}{4\pi} = \frac{4 r_H - \frac{2 M}{r_H^3}}{4 \pi},\no\\
	&\mu = \lim\limits_{r \to \infty} a_t(r) = Q.
\end{align} 
An interesting issue about this background is that the thermal stability condition, i.e. $T > 0$ imposes a certain bound on available charge and mass
\begin{align}\label{con}
	0 < Q < r_H\sqrt{6}, \qquad 0 < M < 2 r_H^4.
\end{align}
From equations \eqref{bv} and \eqref{soladsrn} we can obtain  the butterfly velocity for AdS-RN as $v_B^2 = \frac{2 \pi T}{3 r_H} = \frac{2}{3} \left(1 - \frac{M}{4 r_H^4}\right)$. Due to the constraint \eqref{con}, $v_B^2 \geq \frac{1}{3}$ for every black hole solution. This observation was also seen in the 1RCBH model.

 In the spin-0 sector of fluctuations on top of  the AdS-RN background, perturbations in the EF coordinates are 
\begin{align}
\delta \Phi = \bigg(\delta g_{v v}, \delta g_{r v}, \delta g_{rr}, \delta g_{r z}, \delta g_{v z}, \delta g_{x x}, \delta g_{z z}, \delta A_{v}, \delta A_{z}\bigg),
\end{align}
By expanding the equations near the horizon, we observe that  the $\mathcal{G}_{v v}$ equation takes the following form
\begin{align}
\frac{\delta g^{0}_{r v}(r_H) \left(k^2- i \frac{k_0^2}{2 \pi T} \omega \right)}{r_H^2}+(\omega -2 i \pi  T) (2 k \delta g^{0}_{z v}(r_H)+2 \omega  \delta g^{0}_{x x}(r_H)+\omega  \delta g^{0}_{z z}(r_H)) = 0,
\end{align}
where $k_0^2 = 6 \pi T r_H$. At the chaos point $\omega = 2 i \pi T$ and $k = i k_0$ this equation becomes a trivial identity and a  family  of different ingoing modes emerge toward this point at different lines with slope
\begin{align}
\frac{\delta \omega}{\delta k} = \frac{2 k_0 \delta g_{v v}^{0}(r_H)}{\frac{k_0^2}{2 \pi T}\delta g_{v v}^{0}(r_H) - 2 k_0 r_H^2 \delta g_{z v}^{0}(r_H) - 2 \pi  T r_H^2 \left(2\delta g_{x x}^{0}(r_H) + \delta g_{z z}^{0}(r_H)\right)}.
\end{align}
As we did in the previous subsection, to decouple the linearized equations  we choose the following two invariant fluctuations
\begin{align}\label{eqrn}
\psi_1 &\equiv \frac{d}{dr}\left[\frac{2\delta g_{x x}(r)  + \delta g_{z z}(r)}{r^2}\right] - \frac{i \omega}{r^2 h(r)} \left(2\delta g_{x x}(r)  + \delta g_{z z}(r)\right) - \frac{2 i k}{r^2} \left(\delta g_{r z}(r) + \frac{1}{h(r)}\delta g_{v z}(r)\right)\no\\
& - \frac{3 h(r)}{r} \left(\delta g_{rr}(r) +  \frac{2}{h(r)}\delta g_{r v}(r) + \frac{1}{h(r)^2} \delta g_{v v}(r)\right) - \left(\frac{k^2}{r^3 h(r)} + \frac{3}{ r^2} \left(\frac{f'(r)}{2 h(r)}  -  \frac{1}{r} \right)\right) \delta g_{x x}(r) ,\no\\
\psi_2 &\equiv k \delta A_v(r) + \omega \delta A_z(r) - \frac{k a_t'(r)}{2 r} \delta g_{x x}(r),
\end{align} 
and keep track of the mentioned steps  to derive the linearized equation for invariant perturbations. The brief result  is reported below
\begin{align}\label{eqinvariantrn}
&f^{(1)}(r) \psi_1(r) + f^{(2)}(r) \psi_2(r)  +g^{(1)}(r) \psi_1'(r) +g^{(2)}(r) \psi_2'(r) +  \psi_2''(r) = 0,\\
&h^{(1)}(r) \psi_1(r) + h^{(2)}(r) \psi_2(r)  + I^{(1)}(r) \psi_1'(r) + I^{(2)}(r) \psi_2'(r) + K^{(1)}(r) \psi_1''(r) + K^{(2)}(r) \psi_2''(r)= 0,\nonumber
\end{align}  
where $f^{(i)} = \frac{\mathcal{N}_f^i}{\mathcal{D}}$, $g^{(i)} = \frac{\mathcal{N}_g^i}{\mathcal{D}}$, $h^{(i)} = \frac{\mathcal{N}_h^i}{\mathcal{D}}$, $ I^{(i)} = \frac{\mathcal{N}_I^i}{\mathcal{D}}$ and $K^{(i)} = \frac{\mathcal{N}_K^i}{\mathcal{D}}$. We write the precise form of the coefficients in a more detailed one,  $\mathcal{N}_f^i$, $\mathcal{N}_g^i$, $\mathcal{N}_h^i$, $\mathcal{N}_I^i$, $\mathcal{N}_K^i$ and $\mathcal{D}$ in  appendix \ref{Appendix2}. To explore the pole-skipping feature in the AdS-RN background, we have to plug the ansatz \eqref{eq419} for each perturbing field near the horizon and obtain the coefficients $\alpha_i$ in such a way that the most diverging term disappears. We observe that for the invariant combinations of equation \eqref{eqrn}, the coefficients of $\psi_2$ become $(\alpha_1, \alpha_2) = (0, 1)$ at regular points and at chaos point  the solutions are all regular, regardless of $\omega$ and $T$. For $\psi_1$ we  obtain not any constraint on $\alpha$s.
\section{Conclusion}\label{sec5}
In this paper, we study the chaotic properties of the 1RCBH model. This 5-dimensional Einstein-Maxwell-Dilaton  model is dual to a 4-dimensional strongly charged coupled field theory enjoying a critical  point in its parameter space. To diagnose these properties, we use the  OTOCs and pole-skipping analysis. In the latter, the chaotic properties of many-body thermal systems are  encoded in energy density two-point functions $G^{R}_{T^{00}T^{00}}(\omega,k)$ and linearized equations of motion. The butterfly velocity $v_B$ and Lyapunov exponent $ \lambda_{L}$, which are naturally extracted from OTOC, have been studied as the parameters containing the information about chaos. On the other hand, we investigate the pole-skipping phenomenon which concerns the analytic behavior of $G^{R}_{T^{00}T^{00}}(\omega,k)$ around the chaos point  in the complex  $(\omega-k)$ plane. We also address the chaotic behaviors for AdS-RN background and find that $v_B^2 \geq c_s^2$ at every point of black hole solutions. This finding is valid for the 1RCBH model as well.
 We list our main results in the following: 
\begin{itemize}
\item We consider a general asymptotically $AdS$ background  whose metric is given by equation \eqref{general}
and  compute OTOC and then derive holographically  explicit expressions for $ \lambda_{L}$ and $v_B$ 
 \begin{align}
 \lambda_{L}=\frac{2\pi}{\beta},\,\,\,\,\,\,\,\,\,\,v_B=\frac{2\sqrt{\pi}(f_1f_2)^{\frac{1}{4}}}{\sqrt{\beta (d-1)f_3'}}\bigg\vert_{r=r_H},
 \end{align}
 where $\beta$ is the inverse of Hawking's temperature. The above result perfectly matches the previously reported  CFT results.  In the 1RCBH model, we read 
 \begin{align}
 v_B^2=\frac{4}{7\mp \sqrt{1-\left(\frac{\mu /T}{\pi /\sqrt{2}}\right)^2}},
 \end{align}
 where $-(+)$ indicates the stable(unstable) blackhole solutions. Interestingly, we find that the butterfly velocity $v_B$ can be used as a probe to see the critical point of the corresponding dual field theory. Furthermore, we  study the behavior of $v_B$ near this critical point and observe that the dynamical critical exponent is equal to $\frac{1}{2}$ which is in complete agreement with the results reported in the literature. Interestingly, we find $v_B^2 \geq c_s^2$ for every point of $0 \leq \frac{\mu}{T} \leq \frac{\pi}{\sqrt{2}}$.
 \item Related to the pole-skipping phenomenon we observe that the metric component $\delta g_{v v}^{0}(r_H)$ decouples from the other components of the $vv$ component of the linearized equations  at special point   $\omega_{*}=i\lambda_{L}=2 \pi Ti$ and reduces to the following form
 \begin{align}
	\delta g_{v v}^{0}(r_H) \left( k^2 - \frac{i k_{0}^2 \omega}{2 \pi T} \right) = 0.
\end{align}
 which is interestingly identical to the one governing the spatial profile of a gravitational shock wave \eqref{Shift}. Hence, one can use the pole-skipping analysis to determine the chaotic properties of the underlying theory including $v_B$ and  $ \lambda_{L}$.  The other point is that at special point $k=i k_{0}$  equation \eqref{eqgvvred}  is identically satisfied which means that this equation does not impose any constraint on the near-horizon expansion components  $\delta g_{v v}^{0}(r_H)$, $\delta g_{z v}^{0}(r_H)$, $\delta g_{x x}^{0}(r_H)$ and $\delta g_{z z}^{0}(r_H)$. In other words,  there exists an extra linearly independent ingoing solution to Einstein’s equations that lead to the pole-skipping in   $G^{R}_{T^{00}T^{00}}(\omega,k)$. To find a better understanding of this phenomenon,  it is observed that if one moves away from the chaos point, i.e. $\omega = 2 \pi T i$ and $k = i k_0$ along the slope  \eqref{eqslope}, then there will be lines in $G^{R}_{T^{00}T^{00}}(\omega,k)$ that passes through that point. As a result,  this slope can be considered as an extra parameter in the near-horizon solution and can be used to adjust the ingoing solution onto different asymptotic solutions at the boundary. We study also the pole-skipping phenomenon through the linearized equations from another point of view. Indeed, we  decouple the linearized equations by choosing the suitable gauge+diff invariant master field variables focusing on the spin 0 sector. Details of derivations for dynamical equations are given. We provide the exact form of the linearized equations of motion in separate files along with this paper. We investigate pole-skipping of the retarded Green's function at higher Matsubara points in the equations of motion for scalar field perturbations. It is observed that regular solutions near the horizon are labelled with only one unknown coefficient and point $k = i k_0$ not to add further constraint on equations. These issues are not seen in the gauge field perturbation equations.
\item We investigate the chaotic behaviors for the AdS-RN background. We obtain $v_B^2 \geq c_s^2$ for every black hole solution. We obtain the dynamical equations for gauge+diff invariant perturbations  given in the  \eqref{eqrn}, by using the general method based on the properties of Einstein equation.  Having the regular solutions near the horizon is guaranteed once the equation \eqref{eqgvvred} is derived. We didn't observe this kind of feature but it could be achieved by choosing appropriate combinations of invariant perturbations.
\end{itemize}  
Our main concern in this study is to check whether the critical models respond to  chaotic conditions.  Several questions  call which we leave for further investigations. One can investigate the OTOC and pole-skipping for the Non-conformal backgrounds having special critical points. Also  examine the chaotic behaviors for the gravitational models having large similarity  to the QCD phase diagram deserve further investigation. These works are postponed to our future works. Another interesting work is to look more carefully  the relation $v_B^2 \geq c_s^2$ by mathematical arguments.
\section*{Acknowledgement}
We would like to thank Saso Grozdanov for valuable discussions and for comments on the draft of this paper. The authors also acknowledge Matteo Baggioli  for fruitful discussion.

\appendix
\section{diff+gauge transformations}\label{Appendix1}
In the spin-zero sector, there are seven combinations of fluctuation which are invariant under the diff+gauge transformations \cite{Jansen:2019wag}. By the metric given in the equation \eqref{general} (in the schwarzschild coordinates) and for the following choice of fluctuations set
\begin{align}
	&\delta g_{\mu \nu}(t, r, z) = e^{ i k z} \delta g_{\mu \nu}(t, r), \no\\
	&\delta A_{\mu }(t, r, z) = e^{ i k z} \delta A_{\mu }(t, r),\no\\
	&\delta \phi(v, r, z) = e^{ i k z} \delta \phi(t, r),
\end{align}
they can be written as follows
\begin{align}\label{invsch}
	\Phi_1^{(Sch)}(r) &= \delta g_{t t}(t, r) + \frac{2 i}{k} \partial_t \delta g_{t z}(t, r) + \partial_t^2 \delta g_-(t, r) + \frac{f_1'(r)}{f_3'(r)} \delta g_{x x}(t, r),\no\\
	\Phi_2^{(Sch)}(r) &= \delta g_{t r}(t, r) + \frac{i}{k} \partial_r \delta g_{t z}(t, r) + \bigg(\frac{1}{2} \partial_r \partial_t - \frac{f_1'(r)}{2 f_1(r)} \partial_t\bigg)\delta g_-(t, r) - \frac{i f_1'(r)}{k f_1(r)} \delta g_{t z}(t, r)\no\\
	&  - \frac{f_2(r)}{f_3'(r)} \partial_t \delta g_{x x}(t, r),\no\\
	\Phi_3^{(Sch)}(r) &= \delta g_{r r}(t, r) - \frac{ f_2(r)}{f_3'(r)} \bigg(2 \partial_r + \frac{f_2'(r)}{f_2(r)}  - 2 \frac{f_3''(r)}{f_3'(r)}\bigg)\delta g_{x x}(t, r),\no\\
	\Phi_4^{(Sch)}(r) &= - \frac{i}{k} \delta g_{r z}(t, r)  +\bigg( \frac{f_3'(r)}{2 f_3(r)} - \frac{1}{2} \partial_r \bigg)\delta g_-(t, r) - \frac{f_2(r)}{f_3'(r)} \delta g_{x x}(t, r),\no\\
	\Phi_5^{(Sch)}(r) &= \delta A_t(t, r) + \frac{i}{k} \partial_t \delta A_z(t, r) - \frac{A_t'(r)}{f_3'(r)} \delta g_{x x}(t, r),\no\\
	\Phi_6^{(Sch)}(r) &= \delta A_r(t, r) + \frac{i}{k} \partial_r \delta A_z(t, r) + \frac{A_t'(r)}{2 f_1(r)} \partial_t \delta g_-(t, r) + \frac{i A_t'(r)}{k f_1(r)} \delta g_{t z}(t, r),\no\\
	\Phi_7^{(Sch)}(r) &= \delta \phi(t, r) - \frac{\phi'(r)}{f_3'(r)} \delta g_{x x}(t, r),
\end{align} 
where $\delta g_- = \frac{\delta g_{x x} - \delta g_{z z}}{q^2}$. To convert these fluctuations to the EF coordinates, we must  notice to the fluctuations and derivatives transformation rules 
\begin{align}
	&\delta g^{(Sch)}_{a b} = \frac{\partial x^\mu}{\partial x^a}\, \frac{\partial x^\nu}{\partial x^b}\delta g^{(EF)}_{\mu \nu},\no\\
	&\partial_a^{(Sch)} = \frac{\partial x^\mu}{\partial x^a}\,\partial_\mu^{(EF)}.
\end{align} 
According to the coordinate relation $t = v - g(r)$ with $g'(r) = \sqrt{\frac{f_2(r)}{f_1(r)}}$, the fluctuations set transformations result to
\begin{align}\label{changefluc}
	&\delta g_{t t}(t, r) = \delta g_{v v}(v, r), \qquad \delta g_{t r}(t, r) = \sqrt{\frac{f_2(r)}{f_1(r)}} \delta g_{v v}(v, r) + \delta g_{v r}(v, r),\no\\
	&\delta g_{r r}(t, r) = \delta g_{r r}(v, r) + 2 \sqrt{\frac{f_2(r)}{f_1(r)}} \delta g_{v r}(v, r) + \frac{f_2(r)}{f_1(r)}\delta g_{v v}(v, r),\no\\
	&\delta g_{t z}(t, r) = \delta g_{v z}(v, r), \qquad \delta g_{r z}(t, r) = \sqrt{\frac{f_2(r)}{f_1(r)}} \delta g_{v z}(v, r) + \delta g_{r z}(v, r),\\
	& \delta g_{i i}(t, r) = \delta g_{i i}(v, r), \,\, i = x, y, z, \qquad \delta \phi(t, r) = \delta \phi(v, r)\no\\
	& \delta A_t(t, r) = \delta A_v(v, r), \quad \delta A_z(t, r) = \delta A_z(v, r), \quad \delta A_r(t, r) = \delta A_r(v, r) + \sqrt{\frac{f_2(r)}{f_1(r)}} \delta A_{v}(v, r).\no
\end{align}
Moreover, derivatives transform as follows
\begin{align}\label{changeder}
	\partial_t \to \partial_v, \qquad \partial_r \to \partial_r + \sqrt{\frac{f_2(r)}{f_1(r)}} \,\partial_v.
\end{align}
The replacements \eqref{changefluc} and \eqref{changeder} have to be inserted into the equations \eqref{invsch} to obtain invariant fluctuations in the EF coordinates. The result is as follows
\begin{align}
	\Phi_1^{(EF)}(r) &= \delta g_{v v}(v, r) + \frac{2 i}{k} \partial_v \delta g_{v z}(v, r) + \partial_v^2 \delta g_-(v, r) + \frac{f_1'(r)}{f_3'(r)} \delta g_{x x}(v, r),\no\\
	\Phi_2^{(EF)}(r) &= \sqrt{\frac{f_2(r)}{f_1(r)}} \delta g_{v v}(v, r) + \delta g_{v r}(v, r) + \frac{i}{k} \bigg(\partial_r + \sqrt{\frac{f_2(r)}{f_1(r)}} \,\partial_v\bigg) \delta g_{v z}(v, r)  \no\\
	& + \bigg(\frac{1}{2}  \partial_v \partial_r + \frac{1}{2} \sqrt{\frac{f_2(r)}{f_1(r)}} \,\partial_v^2- \frac{f_1'(r)}{2 f_1(r)} \partial_v\bigg)\delta g_-(v, r) - \frac{i f_1'(r)}{k f_1(r)} \delta g_{v z}(v, r) - \frac{f_2(r)}{f_3'(r)} \partial_v \delta g_{x x}(v, r),\no\\
	\Phi_3^{(EF)}(r) &=  \delta g_{r r}(v, r) + 2 \sqrt{\frac{f_2(r)}{f_1(r)}} \delta g_{v r}(v, r) + \frac{f_2(r)}{f_1(r)}\delta g_{v v}(v, r)\no\\
	& - \frac{f_2(r)}{f_3'(r)} \bigg(2 \partial_r + 2 \sqrt{\frac{f_2(r)}{f_1(r)}} \,\partial_v + \frac{f_2'(r)}{f_2(r)}  - 2 \frac{f_3''(r)}{f_3'(r)}\bigg)\delta g_{x x}(v, r),\no\\
	\Phi_4^{(EF)}(r) &= - \frac{i}{k} \bigg(\sqrt{\frac{f_2(r)}{f_1(r)}} \delta g_{v z}(v, r) + \delta g_{r z}(v, r)\bigg)  +\bigg( \frac{f_3'(r)}{2 f_3(r)} - \frac{1}{2} \partial_r -\frac{1}{2} \sqrt{\frac{f_2(r)}{f_1(r)}} \,\partial_v \bigg)\delta g_-(v, r) \no\\
	&- \frac{f_2(r)}{f_3'(r)} \delta g_{x x}(v, r),\no\\
	\Phi_5^{(EF)}(r) &= \delta A_v(v, r) + \frac{i}{k} \partial_v \delta A_z(v, r) - \frac{A_t'(r)}{f_3'(r)} \delta g_{x x}(v, r),\no\\
	\Phi_6^{(EF)}(r) &= \delta A_r(v, r)  + \sqrt{\frac{f_2(r)}{f_1(r)}} \delta A_{v}(v, r) +\frac{i}{k} \bigg(\partial_r + \sqrt{\frac{f_2(r)}{f_1(r)}} \,\partial_v\bigg) \delta A_z(v, r) \no\\
	&+ \frac{A_t'(r)}{2 f_1(r)} \partial_v \delta g_-(v, r) + \frac{i A_t'(r)}{k f_1(r)} \delta g_{v z}(v, r),\no\\
	\Phi_7^{(EF)}(r) &= \delta \phi(v, r) - \frac{\phi'(r)}{f_3'(r)} \delta g_{x x}(t, r),
\end{align}
It is obvious that any combination of these invariant functions is again an invariant choice. For instance, the  combinations of the equation \eqref{eq47} are nothing but 
\begin{align}
	&\psi_1 = \frac{2 k^2}{f_3(r)} \Phi_4^{(EF)}(r) - \frac{3 f_3'(r)}{2 f_2(r) f_3(r)} \Phi_3^{(EF)}(r),\no\\
	&\psi_2 = \Phi_7^{(EF)}(r), \qquad \psi_3 = \Phi_5^{(EF)}(r).
\end{align}
\section{Details of the near horizon expansion}\label{Appendix3}
In this part, we present the details of coefficients $M_{i j}(\omega, k^2)$ in the equation \eqref{pole}. The first elements are
\begin{align}
	&M_{1 1}(\omega, k^2) = \frac{1}{6} \left(e^{-\frac{1}{2} \sqrt{\frac{3}{2}} \phi(r_H)} \left(-3 k^2 e^{\sqrt{\frac{2}{3}} \phi(r_H)-2 A(r_H)}+68 e^{\sqrt{\frac{3}{2}} \phi(r_H)}+40\right)-3 (32 \pi  T+3 i \omega ) A'(r_H)\right),\nonumber\\
	&M_{2 1}(\omega, k^2) = \frac{i k_0^4 \omega  e^{-4 A(r_H)+\frac{\phi(r_H)}{2 \sqrt{6}}}}{24 \pi ^2 T^2} + \frac{k_0^2 e^{-4 A(r_H)-\frac{3}{2} \sqrt{\frac{3}{2}} \phi(r_H)}}{144 \pi ^2 T^2} \no\\
	& \times \left(-3 i \pi  T \omega  e^{2 A(r_H)+\frac{5 \phi(r_H)}{\sqrt{6}}} \left(12 A'(r_H)-\sqrt{6} \phi'(r_H)\right)-32 \pi  T e^{2 A(r_H)+\frac{\phi(r_H)}{2 \sqrt{6}}} a_t'(r_H)^2+24 \pi  k^2 T e^{\frac{11 \phi(r_H)}{2 \sqrt{6}}}\right)\no\\
	&+\frac{1}{18} e^{\frac{-3}{2} \sqrt{\frac{3}{2}} \phi(r_H)} \bigg(3 \left(8 a_t'(r_H) a_t''(r_H)-9 i \omega  e^{\frac{3}{2} \sqrt{\frac{3}{2}} \phi(r_H)} A''(r_H)\right)+24 A'(r_H) a_t'(r_H)^2\no\\
	&+2 \sqrt{6} \phi'(r_H) \left(-5 a_t'(r_H)^2+e^{\sqrt{6} \phi(r_H)}-4 e^{\sqrt{\frac{3}{2}} \phi(r_H)}\right)\bigg),\nonumber\\
	&M_{2 2}(\omega, k^2) = \frac{e^{\frac{\phi(r_H)}{2 \sqrt{6}}-2 A(r_H)} k_0^2}{12}  \left(16 -\frac{i \omega }{\pi  T}\right) + \frac{1}{12} \bigg(\left(12 A'(r_H) - \sqrt{6}  \phi'(r_H) \right) (6 \pi  T-i \omega ) -6 k^2 e^{\frac{\phi(r_H)}{2 \sqrt{6}}-2 A(r_H)} \no\\
	&+6 e^{2 A(r_H)-\frac{\phi(r_H)}{2 \sqrt{6}}} h''(r_H)+8 e^{\frac{-3}{2} \sqrt{\frac{3}{2}} \phi(r_H)} a_t'(r_H)^2  +8 e^{\frac{1}{2} \sqrt{\frac{3}{2}} \phi(r_H)}+16 e^{-\frac{1}{2} \sqrt{\frac{3}{2}} \phi(r_H)}\bigg),\nonumber\\
		&M_{3 1}(\omega, k^2) =-\frac{i e^{\frac{\sqrt{6}}{4} \phi(r_H)-6 A(r_H)} \omega  k_0^6}{144 \pi ^3 T^3}+\frac{e^{\frac{-3}{4}  \left(8 A(r_H)+\sqrt{6} \phi(r_H)\right)} k_0^4 }{864 \pi ^3 T^3}\no\\
		&\times \bigg(-48 e^{\sqrt{6} \phi(r_H)} \pi  T k^2+64 e^{2 A(r_H)+\frac{\phi(r_H)}{\sqrt{6}}} \pi  T a_t'(r_H)^2+\left(12 A'(r_H) - \sqrt{6}  \phi'(r_H) \right)  e^{2 A(r_H)+\frac{11\phi(r_H)}{2\sqrt{6}}} i \pi  T \omega  \bigg)\no\\
		&+\frac{e^{\frac{-3}{4}  \left(8 A(r_H)+\sqrt{6} \phi(r_H)\right)} k_0^2 }{864 \pi ^3 T} \bigg(- i \pi ^2 \omega e^{4 A(r_H)+\frac{5 \phi(r_H)}{\sqrt{6}}}  \left(216 A'(r_H)^2 + 9 \phi'(r_H)^2 \right)\no\\
		&  e^{4 A(r_H)+\frac{\phi(r_H)}{2 \sqrt{6}}} \pi ^2 a_t'(r_H)^2 \left(-768 A'(r_H) + 320 \sqrt{6}\phi'(r_H) \right) +e^{4 A(r_H)+ i \pi ^2 \omega  \frac{5 \phi(r_H)}{\sqrt{6}}}  \left(432 A''(r_H) + 18 \sqrt{6} \phi''(r_H)\right)\no\\
		&+36 e^{4 A(r_H)+\frac{11\phi(r_H)}{2 \sqrt{6}}} i \pi ^2 \omega  \sqrt{6} A'(r_H) \phi'(r_H)  -768 e^{4 A(r_H)+\frac{\phi(r_H)}{2 \sqrt{6}}} \pi ^2 a_t'(r_H) a_t''(r_H)  \bigg) +  k^2 e^{\frac{\phi(r_H)}{2 \sqrt{6}}-2 A(r_H)} A''(r_H) \no\\
		&+\frac{ e^{\frac{-3}{2}  \sqrt{\frac{3}{2}} \phi(r_H)}}{9} \bigg(a_t'(r_H)^2 \left(-20 \sqrt{6} A'(r_H) \phi'(r_H)+24 A'(r_H)^2-5 \sqrt{6} \phi''(r_H)+25 \phi'(r_H)^2\right)\no\\
		&\hspace{-1cm}+4 a_t'(r_H) \left(a_t''(r_H) \left(12 A'(r_H)-5 \sqrt{6} \phi'(r_H)\right)+3 a_t^{(3)}(r_H)\right)+12 a_t''(r_H)^2\bigg) +\frac{4 e^{-\frac{1}{2} \sqrt{\frac{3}{2}} \phi(r_H)}}{9} \left(2 \phi'(r_H)^2-\sqrt{6} \phi''(r_H)\right)\no\\
		&- \frac{i \omega}{4}  \left(6 A^{(3)}(r_H)+A''(r_H) \left(12 A'(r_H)-\sqrt{6} \phi'(r_H)\right)\right)+\frac{ e^{\frac{1}{2} \sqrt{\frac{3}{2}} \phi(r_H)}}{9} \left(\sqrt{6} \phi''(r_H)+\phi'(r_H)^2\right),\nonumber
		\end{align}
		\begin{align}
			&M_{3 2}(\omega, k^2) = \frac{i k_0^4 \omega  e^{\frac{\phi(r_H)}{\sqrt{6}}-4 A(r_H)}}{18 \pi ^2 T^2} + \frac{1}{72} k_0^2 \bigg(\frac{12 (32 \pi  T-i \omega ) e^{\frac{\phi(r_H)}{2 \sqrt{6}}-2 A(r_H)} A'(r_H)}{\pi  T} +\frac{i \sqrt{6} \omega  e^{\frac{\phi(r_H)}{2 \sqrt{6}}-2 A(r_H)} \phi'(r_H)}{\pi  T}\no\\
			&-\frac{32 e^{-2 A(r_H)-2 \sqrt{\frac{2}{3}} \phi(r_H)} a_t'(r_H)^2}{\pi  T}+\frac{24 q^2 e^{\frac{\phi(r_H)}{\sqrt{6}}-4 A(r_H)}}{\pi  T}-32 \sqrt{6} e^{\frac{\phi(r_H)}{2 \sqrt{6}}-2 A(r_H)} \phi'(r_H)+\frac{24 h''(r_H)}{\pi  T}\bigg) \no\\
			&+ \frac{1}{72} \bigg(216 (8 \pi  T-i \omega ) A''(r_H)+12 A'(r_H) \bigg(15 e^{2 A(r_H)-\frac{\phi(r_H)}{2 \sqrt{6}}} h''(r_H)+16 e^{\frac{-3}{2} \sqrt{\frac{3}{2}} \phi(r_H)} a_t'(r_H)^2\no\\
			&+\sqrt{6} (-16 \pi  T+i \omega ) \phi'(r_H)\bigg)+72 (16 \pi  T-i \omega ) A'(r_H)^2 +36 h^{(3)}(r_H) e^{2 A(r_H)-\frac{\phi(r_H)}{2 \sqrt{6}}}\no\\
			&+\sqrt{6} e^{\frac{-3}{2} \sqrt{\frac{3}{2}} \phi(r_H)} \phi'(r_H) \left(-15 e^{2 A(r_H)+2 \sqrt{\frac{2}{3}} \phi(r_H)} h''(r_H)-80 a_t'(r_H)^2+16 e^{\sqrt{6} \phi(r_H)}-64 e^{\sqrt{\frac{3}{2}} \phi(r_H)}\right)\no\\
			&+192 e^{\frac{-3}{2}  \sqrt{\frac{3}{2}} \phi(r_H)} a_t'(r_H) a_t''(r_H)-48 \sqrt{6} \pi  T \phi''(r_H)+3 (16 \pi  T-i \omega ) \phi'(r_H)^2+6 i \sqrt{6} \omega  \phi''(r_H)\bigg),\nonumber\\
				&M_{3 3}(\omega, k^2) = \frac{1}{12} k_0^2 e^{\frac{\phi(r_H)}{2 \sqrt{6}}-2 A(r_H)} \left(32 +\frac{i \omega }{\pi  T}\right)  + 2  (10 \pi  T-i \omega ) A'(r_H) + \frac{3}{2}  e^{2 A(r_H)-\frac{\phi(r_H)}{2 \sqrt{6}}} h''(r_H)\no\\
				&\hspace{-1.5cm}+ \frac{1}{12} \bigg(-6 k^2 e^{\frac{\phi(r_H)}{2 \sqrt{6}}-2 A(r_H)}+8 e^{\frac{-3}{2}  \sqrt{\frac{3}{2}} \phi(r_H)} a_t'(r_H)^2-20 \sqrt{6} \pi  T \phi'(r_H)+2 i \sqrt{6} \omega  \phi'(r_H)+8 e^{\frac{1}{2} \sqrt{\frac{3}{2}} \phi(r_H)}+16 e^{-\frac{1}{2} \sqrt{\frac{3}{2}} \phi(r_H)}\bigg).\nonumber
\end{align}
\section{Coefficients for AdS-RN linearized equations}\label{Appendix2}
 Here we list the coefficients of the first and the second equation of \eqref{eqinvariantrn}
\begin{align}
	&\hspace{-1cm}\mathcal{N}_g^1 =  k r h(r) a_t'(r) \bigg\{4 r^2 a_t'(r)^2 \left(9 k^2 h(r)+k^4+3 r^2 \omega ^2\right)-k^2 r^4 a_t'(r)^4-48 h(r) \left(6 k^2 h(r)+k^4+6 k^2 r^2-6 r^2 \omega ^2\right)\no\\
	&+48 r^2 \left(k^2+6 r^2\right) \left(2 k^2-\omega ^2\right)\bigg\},\no\\
		&\hspace{-1cm}\mathcal{D} =  -3 r^4 a_t'(r)^4 \left(3 k^2 h(r)+r^2 \omega ^2\right)+24 r^2 a_t'(r)^2 \bigg(h(r) \left(k^4+6 k^2 r^2-6 r^2 \omega ^2\right)+r^2 \omega ^2 \left(k^2+6 r^2\right)\bigg)\no\\
		&+48 \left(-6 h(r)+k^2+6 r^2\right)^2 \left(k^2 h(r)-r^2 \omega ^2\right) ,\no\\
		&\hspace{-1cm}\mathcal{N}_g^2  = \frac{1}{2 r h(r)} \bigg\{-576 h(r)^3 \left(k^2 r^2 a_t'(r)^2+6 \left(k^4+2 k^2 r (3 r+i \omega )-r^2 \omega ^2\right)\right)\no\\
		&+2 r h(r) \left(r^2 a_t'(r)^2-4 \left(k^2+6 r^2\right)\right) \bigg(k^2 r^3 a_t'(r)^4+r a_t'(r)^2 \left(4 k^4+6 k^2 r (-4 r+3 i \omega )+27 r^2 \omega ^2\right)\no\\
		&+12 i \omega  \left(2 k^4+k^2 r (12 r+i \omega )+6 r^2 \omega  (4 \omega +9 i r)\right)\bigg)+6 h(r)^2 \bigg(-5 k^2 r^4 a_t'(r)^4\no\\
		&\hspace{-0.5cm}+8 a_t'(r)^2 \left(5 k^4 r^2+6 k^2 r^4+18 r^4 \omega ^2\right)+48 \left(\left(k^3+6 k r^2\right)^2-4 r^2 \omega ^2 \left(k^2+18 r^2\right)+8 i k^2 r \omega  \left(k^2+6 r^2\right)+24 i r^3 \omega ^3\right)\bigg)\no\\
		&+r^3 \omega ^2 \left(r^2 a_t'(r)^2-4 \left(k^2+6 r^2\right)\right)^2 \left(r a_t'(r)^2-24 r+12 i \omega \right)+10368 k^2 h(r)^4\bigg\},\no\\
	&\hspace{-1cm}\mathcal{N}_f^1  =  -a_t'(r) \bigg\{288 k^3 r h(r)^3 - k r^2 h(r)  \bigg(-5 k^2 r^3 a_t'(r)^4+4 r a_t'(r)^2 \left(7 k^4+9 k^2 r (4 r-i \omega )+21 r^2 \omega ^2\right)\no\\
	&+48 \left(k^4 (2 r+i \omega )+3 k^2 r \left(-4 r^2+2 i r \omega +\omega ^2\right)+6 r^2 \omega ^2 (11 r-i \omega )\right)\bigg)\no\\
	&+12 h(r)^2 \left(-5 k^3 r^3 a_t'(r)^2+4 k r  \left(5 k^4+6 k^2 r (5 r-i \omega )+18 r^2 \omega ^2\right)\right)\no\\
	&\hspace{-0.5cm}-k r^4 \left(r^2 a_t'(r)^2-4 \left(k^2+6 r^2\right)\right) \bigg( \omega  a_t'(r)^2 \left(2 r \omega +i k^2\right)+\frac{k^2}{6} r a_t'(r)^4-12 \left(k^2 (8 r-2 i \omega )+\omega ^2 (-8 r+i \omega )\right)\bigg)\bigg\},\no\\
	&\hspace{-1cm}\mathcal{N}_f^2  = -\frac{1}{2 r^2 h(r)}\bigg\{3456 h(r)^3 \left(k^4+5 i k^2 r \omega \right) -144 h(r)^2 \bigg(4 r^2 a_t'(r)^2 \left(k^4+6 r^2 \omega ^2\right)-k^2 r^4 a_t'(r)^4\no\\
	&+8 \left(k^6+k^4 r (6 r+5 i \omega )+3 k^2 r^2 \omega  (\omega +14 i r)+3 i r^3 \omega ^3\right)\bigg)\no\\
	&\hspace{-0.5cm}+6 h(r) \left(4 \left(k^2+6 r^2\right)-r^2 a_t'(r)^2\right) \bigg(2 k^2 r^4 a_t'(r)^4+r^2 a_t'(r)^2 \left(7 k^4+11 i k^2 r \omega +48 r^2 \omega ^2\right)\no\\
	&+4 \left(k^6+k^4 r (6 r+5 i \omega )+6 k^2 r^2 \omega  (2 \omega +13 i r)+12 i r^3 \omega ^3\right)\bigg)\no\\
	&\hspace{-0.5cm}-i \omega  \left(r^3 a_t'(r)^2-4 r \left(k^2+6 r^2\right)\right)^2 \left(r a_t'(r)^2 \left(k^2-6 i r \omega \right)+6 k^2 (4 r-i \omega )+6 r \omega ^2\right)\bigg\}.
\end{align}
\begin{align}
	&\hspace{-1cm}\mathcal{N}_K^1 = \frac{r h(r)}{2} \bigg\{-48 h(r)^2 \bigg(k^2 r^2 a_t'(r)^2+8 k^4+12 k^2 r (4 r+i \omega )+36 r^2 \omega ^2\bigg)\no\\
	&+h(r) \bigg(-3 k^2 r^4 a_t'(r)^4+8 r^2 \left(k^4+6 k^2 r (r+i \omega )-12 r^2 \omega ^2\right) a_t'(r)^2\no\\
	&+16 \left(k^6+12 k^4 r^2+12 k^2 \left(3 r^4+2 r^2 \omega ^2\right)+36 r^3 \omega ^2 (4 r+i \omega )\right)\bigg)\no\\
	&-r^2 \omega ^2 \left(r^2 a_t'(r)^2-4 \left(k^2+6 r^2\right)\right)^2+1728 k^2 h(r)^3\bigg\}\no\\
	&\hspace{-1cm}\mathcal{N}_K^2 = 12 k r h(r) a_t'(r) \bigg(h(r) \left(2 r^2 a_t'(r){}^2-8 \left(k^2+6 r^2\right)\right)+i r \omega  \left(r^2 a_t'(r){}^2-4 \left(k^2+6 r^2\right)\right)+48 h(r)^2\bigg),\no\\
		&\hspace{-1cm}\mathcal{N}_I^1 = \frac{1}{12} \bigg\{17915904 k^4 h(r)^7-165888 h(r)^6 \bigg(7 r^2 a_t'(r)^2 k^4+36 \left(k^6+r (3 i \omega -2 r) k^4+6 r^2 \omega ^2 k^2\right)\bigg) \no\\
		&+13824 h(r)^5 \bigg(-17 k^4 r^4 a_t'(r)^4+4 k^2 r^2 \left(16 k^4+15 r (6 r+i \omega ) k^2-24 r^2 \omega ^2\right) a_t'(r)^2\no\\
		&+16 \left(2 k^8+3 r (11 i \omega -38 r) k^6-18 r^2 \left(42 r^2-7 i \omega  r-2 \omega ^2\right) k^4-54 r^3 (2 r-3 i \omega ) \omega ^2 k^2+81 r^4 \omega ^4\right)\bigg) \no\\
		&+288 h(r)^4 \bigg(71 k^4 r^6 a_t'(r)^6-12 k^2 r^4 \left(3 k^4+9 r (14 r-5 i \omega ) k^2-136 r^2 \omega ^2\right) a_t'(r)^4\no\\
		&-48 r^2 a_t'(r)^2 \left(13 k^8+2 r (38 r+21 i \omega ) k^6-12 r^2 \left(r^2-15 i \omega  r-4 \omega ^2\right) k^4+96 r^3 \omega ^2 (10 r-i \omega ) k^2-180 r^4 \omega ^4\right) \no\\
		&\hspace{-0.5cm}+64 \bigg(k^{10}+3 r (130 r-23 i \omega ) k^8+12 r^2 \left(381 r^2-39 i \omega  r+2 \omega ^2\right) k^6+36 r^3 \left(378 r^3-9 i \omega  r^2+88 \omega ^2 r-22 i \omega ^3\right) k^4\no\\
		&+108 r^4 \omega ^2 \left(168 r^2-28 i \omega  r+\omega ^2\right) k^2+324 r^5 \omega ^4 (2 r-3 i \omega )\bigg)\bigg) -6 h(r)^3 \bigg(-3 k^4 r^8 a_t'(r)^8\no\\
		&+16 k^2 r^6 \left(34 k^4+3 r (62 r+39 i \omega ) k^2-90 r^2 \omega ^2\right) a_t'(r)^6-96 k^2 r^4 a_t'(r)^4 \bigg(17 k^6+16 r (23 r+i \omega ) k^4\no\\
		&+4 r^2 \left(399 r^2+66 i \omega  r-46 \omega ^2\right) k^2-36 r^3 (26 r+11 i \omega ) \omega ^2\bigg) -256 r^2  a_t'(r)^2\bigg(8 k^{10}+3 r (17 i \omega -6 r) k^8\no\\
		&-6 r^2 \left(180 r^2-30 i \omega  r-23 \omega ^2\right) k^6-36 r^3 \left(114 r^3+21 i \omega  r^2-110 \omega ^2 r-i \omega ^3\right) k^4\no\\
		&+72 r^4 \omega ^2 \left(261 r^2+9 i \omega  r-\omega ^2\right) k^2+108 r^5 \omega ^4 (42 r-13 i \omega )\bigg)+256 \bigg(k^{12}+72 r (6 r-i \omega ) k^{10}\no\\
		&+24 r^2 \left(315 r^2-36 i \omega  r+4 \omega ^2\right) k^8+72 r^3 \left(624 r^3-36 i \omega  r^2+142 \omega ^2 r-23 i \omega ^3\right) k^6\no\\
		&+144 r^4 \left(621 r^4+780 \omega ^2 r^2-78 i \omega ^3 r+10 \omega ^4\right) k^4+432 r^5 \omega ^2 \left(756 r^3-18 i \omega  r^2+62 \omega ^2 r-11 i \omega ^3\right) k^2\no\\
		&+2592 r^6 \omega ^4 \left(42 r^2-7 i \omega  r+\omega ^2\right)\bigg)\bigg) -r h(r)^2 \left(4 \left(k^2+6 r^2\right)-r^2 a_t'(r){}^2\right) \no\\
		&\times \bigg(-21 k^4 r^7 a_t'(r)^8+4 k^2 r^5 \left(11 k^4+3 r (52 r+15 i \omega ) k^2-207 r^2 \omega ^2\right) a_t'(r)^6\no\\
		&\hspace{-0.5cm}+48 r^3 \left(3 k^8+r (34 r+11 i \omega ) k^6+r^2 \left(96 r^2-78 i \omega  r-11 \omega ^2\right) k^4+6 r^3 \omega ^2 (105 r-i \omega ) k^2-198 r^4 \omega ^4\right) a_t'(r)^4\no\\
		&+64 r  a_t'(r)^2 \bigg(k^{10}+3 r (5 i \omega -16 r) k^8+9 r^2 \left(-76 r^2+4 i \omega  r+11 \omega ^2\right) k^6\no\\
		&\hspace{-0.5cm}-18 r^3 \left(120 r^3+18 i \omega  r^2-66 \omega ^2 r-13 i \omega ^3\right) k^4+36 r^4 \omega ^2 \left(99 r^2-33 i \omega  r-\omega ^2\right) k^2+162 r^5 \omega ^4 (64 r-5 i \omega )\bigg)\no
		\end{align}
		\begin{align}
		&-768 \bigg((6 r-i \omega ) k^{10}+r \left(108 r^2-18 i \omega  r+\omega ^2\right) k^8+6 r^2 \left(108 r^3-18 i \omega  r^2+71 \omega ^2 r-12 i \omega ^3\right) k^6\no\\
		&+6 r^3 \left(216 r^4-36 i \omega  r^3+834 \omega ^2 r^2-72 i \omega ^3 r+13 \omega ^4\right) k^4+18 r^4 \omega ^2 \left(828 r^3+152 \omega ^2 r-23 i \omega ^3\right) k^2\no\\
		&+108 r^5 \omega ^4 \left(126 r^2-3 i \omega  r+4 \omega ^2\right)\bigg)\bigg) +2 r^3 \omega ^2 h(r) \left(r^2 a_t'(r)^2-4 \left(k^2+6 r^2\right)\right)^2 \no\\
		&\times \bigg(-8 k^2 r^5 a_t'(r)^6+3 r^3 \left(8 k^4+4 r (26 r+3 i \omega ) k^2-57 r^2 \omega ^2\right) a_t'(r)^4\no\\
		&+8 r a_t'(r)^2 \left(4 k^6+12 i r \omega  k^4-3 r^2 \left(48 r^2+48 i \omega  r-17 \omega ^2\right) k^2+1134 r^4 \omega ^2\right) \no\\
		&-48 \bigg(4 (6 r-i \omega ) k^6+r \left(288 r^2-48 i \omega  r+\omega ^2\right) k^4+12 r^2 k^2\left(72 r^3-12 i \omega  r^2+35 \omega ^2 r-6 i \omega ^3\right) \no\\
		&+36 r^3 \omega ^2 \left(69 r^2+2 \omega ^2\right)\bigg)\bigg) -3 r^5 \omega ^4 \left(r a_t'(r)^2-24 r+4 i \omega \right) \left(r^2 a_t'(r)^2-4 \left(k^2+6 r^2\right)\right)^4\bigg\}\no\\
		&\hspace{-1cm}\mathcal{N}_I^2 = - \frac{k a_t'(r)}{12} \bigg\{165888 h(r)^5\left(7 k^2 r^2 a_t'(r)^2+4 \left(k^4+6 r (7 r-i \omega ) k^2+24 r^2 \omega ^2\right)\right)\no\\
		& +6912  h(r)^4 \bigg(-7 k^2 r^4 a_t'(r)^4-12 r^2 \left(4 k^4+3 r (4 r+3 i \omega ) k^2-10 r^2 \omega ^2\right) a_t'(r)^2\no\\
		&+48 \left(k^6+5 r (i \omega -8 r) k^4-2 r^2 \left(138 r^2-3 i \omega  r+13 \omega ^2\right) k^2+12 r^3 \omega ^2 (i \omega -23 r)\right)\bigg)\no\\
		&-288 h(r)^3 \bigg(11 k^2 r^6 a_t'(r)^6-4 r^4 \left(19 k^4+6 r (7 i \omega -11 r) k^2+30 r^2 \omega ^2\right) a_t'(r)^4\no\\
		&\hspace{-0.5cm}-48 r^2 \left(k^6+2 r (5 i \omega -2 r) k^4-4 r^2 \left(15 r^2-9 i \omega  r+4 \omega ^2\right) k^2+6 r^3 \omega ^2 (17 i \omega -104 r)\right) a_t'(r)^2\no\\
		&\hspace{-0.5cm}+64 \bigg(11 k^8-6 r (5 r-4 i \omega ) k^6-18 r^2 \left(86 r^2-4 i \omega  r+\omega ^2\right) k^4-18 r^3 \left(324 r^3+24 i \omega  r^2+100 \omega ^2 r-9 i \omega ^3\right) k^2\no\\
		&+108 r^5 \omega ^2 (5 i \omega -94 r)\bigg)\bigg) +12 h(r)^2 \left(4 \left(k^2+6 r^2\right)-r^2 a_t'(r){}^2\right) \no\\
		&\times \bigg(-9 k^2 r^6 a_t'(r){}^6-4 r^4 \left(13 k^4+3 r (15 i \omega -58 r) k^2+78 r^2 \omega ^2\right) a_t'(r)^4\no\\
		&+16 r^2 \left(13 k^6+6 r (6 r+i \omega ) k^4-36 r^2 \left(7 r^2+3 i \omega  r-\omega ^2\right) k^2+72 r^3 \omega ^2 (28 r-5 i \omega )\right) a_t'(r)^2\no\\
		&+192 \bigg(3 k^8+5 r (2 r+i \omega ) k^6+2 r^2 \left(-102 r^2+6 i \omega  r+19 \omega ^2\right) k^4-12 r^3 \left(78 r^3+9 i \omega  r^2+24 \omega ^2 r-8 i \omega ^3\right) k^2\no\\
		&+72 r^4 \omega ^2 \left(-43 r^2+4 i \omega  r+\omega ^2\right)\bigg)\bigg) -h(r) \left(r^2 a_t'(r)^2-4 \left(k^2+6 r^2\right)\right)^2 \no\\
		&\times \bigg(-3 k^2 r^6 a_t'(r){}^6-4 r^4 \left(k^4-30 r^2 k^2+33 r^2 \omega ^2\right) a_t'(r)^4\no\\
		&+48 r^2 \left(k^6+4 r^2 k^4-4 \left(3 r^4-5 r^2 \omega ^2\right) k^2+3 r^3 \omega ^2 (56 r-9 i \omega )\right) a_t'(r)^2\no\\
		&+64 \bigg(k^8+6 r^2 k^6-9 \left(4 r^4-5 r^2 \omega ^2\right) k^4-9 r^3 \left(24 r^3+4 \omega ^2 r-9 i \omega ^3\right) k^2\no\\
		&-54 r^4 \omega ^2 \left(34 r^2-9 i \omega  r-4 \omega ^2\right)\bigg)\bigg) +r^2 \omega ^2 \left(4 \left(k^2+6 r^2\right)-r^2 a_t'(r)^2\right)^3 \no\\
		&\times \bigg(-r^4 a_t'(r){}^4+24 r^3 (2 r-i \omega ) a_t'(r){}^2+16 \left(k^4+18 r^2 \left(-2 r^2+2 i \omega  r+\omega ^2\right)\right)\bigg)\bigg\},\no
		\end{align}
		\begin{align}
		&\hspace{-1cm}\mathcal{N}_h^1 = -\frac{1}{72 r h(r)} \bigg\{35831808 k^4 h(r)^8+995328 \left(r^2 a_t'(r){}^2 k^4+8 \left(2 k^6+3 r (37 r-8 i \omega ) k^4-9 r^2 \omega ^2 k^2\right)\right) h(r)^7\no\\
		&-1990656 h(r)^6\bigg(k^4 r^4 a_t'(r)^4+k^2 r^3 \left((2 r-i \omega ) k^2+8 r \omega ^2\right) a_t'(r)^2+2 \bigg(k^8+r (128 r-17 i \omega ) k^6\no\\
		&+\left(660 r^4+23 \omega ^2 r^2\right) k^4+12 r^3 \omega ^2 (37 r-8 i \omega ) k^2-9 r^4 \omega ^4\bigg)\bigg) -1728 h(r)^5 \bigg(-103 k^4 r^6 a_t'(r){}^6\no\\
		&+4 k^2 r^4 a_t'(r)^4 \left(7 k^4+33 r (30 r-13 i \omega ) k^2-312 r^2 \omega ^2\right) +16 r^2 a_t'(r)^2 \bigg(11 k^8-6 r (98 r-27 i \omega ) k^6\no\\
		&-36 r^2 \left(121 r^2-13 i \omega  r-3 \omega ^2\right) k^4+144 r^3 \omega ^2 (41 r-7 i \omega ) k^2+36 r^4 \omega ^4\bigg) \no\\
		&+192 \bigg(3 k^{10}+r (33 i \omega -238 r) k^8-12 r^2 \left(213 r^2+17 i \omega  r+5 \omega ^2\right) k^6\no\\
		&-12 r^3 \left(510 r^3+177 i \omega  r^2+262 \omega ^2 r-34 i \omega ^3\right) k^4-24 \left(660 \omega ^2 r^6+17 \omega ^4 r^4\right) k^2+72 r^5 \omega ^4 (8 i \omega -37 r)\bigg)\bigg)\no\\
		& +36 h(r)^4 \bigg(-67 k^4 r^8 a_t'(r)^8-32 k^2 r^6 \left(5 k^4+12 r (14 i \omega -3 r) k^2-153 r^2 \omega ^2\right) a_t'(r)^6\no\\
		&+32 r^4 a_t'(r)^4 \bigg(25 k^8+12 r (123 r-2 i \omega ) k^6+18 r^2 \left(322 r^2+146 i \omega  r+13 \omega ^2\right) k^4+36 r^3 \omega ^2 (106 r-83 i \omega ) k^2\no\\
		&+360 r^4 \omega ^4\bigg) +1024 r^2 \bigg(2 k^{10}+\left(30 i r \omega -87 r^2\right) k^8-9 r^2 \left(148 r^2+3 i \omega  r+2 \omega ^2\right) k^6\no\\
		&\hspace{-0.5cm}-27 r^3 \left(164 r^3+50 i \omega  r^2-41 \omega ^2 r-i \omega ^3\right) k^4+18 r^4 \omega ^2 \left(267 r^2+87 i \omega  r+17 \omega ^2\right) k^2+54 r^5 \omega ^4 (104 r-15 i \omega )\bigg) a_t'(r)^2\no\\
		&+256 \bigg(25 k^{12}+24 r (4 i \omega -13 r) k^{10}-36 r^2 \left(154 r^2+54 i \omega  r-5 \omega ^2\right) k^8\no\\
		&-72 r^3 \left(156 r^3+396 i \omega  r^2+208 \omega ^2 r-33 i \omega ^3\right) k^6+432 r^4 \left(75 r^4-186 i \omega  r^3-393 \omega ^2 r^2-34 i \omega ^3 r-7 \omega ^4\right) k^4\no\\
		&-432 r^5 \omega ^2 \left(1020 r^3+354 i \omega  r^2+140 \omega ^2 r-17 i \omega ^3\right) k^2-6480 r^6 \omega ^4 \left(44 r^2+\omega ^2\right)\bigg)\bigg)\no\\
		& -6 h(r)^3 \bigg(21 k^4 r^{10} a_t'(r)^{10}-2 k^2 r^8 \left(77 k^4+3 r (364 r+51 i \omega ) k^2-690 r^2 \omega ^2\right) a_t'(r)^8\no\\
		&+32 r^6 a_t'(r)^6 \left(10 k^8+9 r (36 r-5 i \omega ) k^6+6 r^2 \left(390 r^2+57 i \omega  r+20 \omega ^2\right) k^4-9 r^3 (502 r-61 i \omega ) \omega ^2 k^2+441 r^4 \omega^4\right)\no\\
		& -64 r^4 a_t'(r)^4\bigg(7 k^{10}+3 r (12 r-41 i \omega ) k^8+12 r^2 \left(159 r^2-126 i \omega  r+22 \omega ^2\right) k^6\no\\
		&\hspace{-0.5cm}+36 r^3 \left(324 r^3-105 i \omega  r^2-170 \omega ^2 r+37 i \omega ^3\right) k^4-54 r^4 \omega ^2 \left(1080 r^2+16 i \omega  r+27 \omega ^2\right) k^2+54 r^5 \omega ^4 (654 r-73 i \omega )\bigg)\no\\
		& \hspace{-0.5cm}+768 r^2 a_t'(r)^2\bigg(k^{12}+2 r (7 i \omega -12 r) k^{10}-12 r^2 \left(56 r^2+3 i \omega  r+3 \omega ^2\right) k^8-6 r^3 \left(768 r^3+324 i \omega  r^2-38 \omega ^2 r-29 i \omega ^3\right) k^6\no\\
		&\hspace{-0.5cm}-6 r^4 \left(1656 r^4+1224 i \omega  r^3+624 \omega ^2 r^2-660 i \omega ^3 r+37 \omega ^4\right) k^4-36 r^5 \omega ^2 \left(1068 r^3-390 i \omega  r^2-354 \omega ^2 r+37 i \omega ^3\right) k^2\no
		\end{align}
		\begin{align}
		&+216 r^6 \omega ^4 \left(363 r^2+29 i \omega  r+7 \omega ^2\right)\bigg) +1536 \bigg(k^{14}+r (4 r+i \omega ) k^{12}+2 r^2 \left(132 r^2-120 i \omega  r+25 \omega ^2\right) k^{10}\no\\
		&+24 r^3 \left(252 r^3-165 i \omega  r^2-14 \omega ^2 r+8 i \omega ^3\right) k^8+36 r^4 \left(1140 r^4-576 i \omega  r^3-212 \omega ^2 r^2-108 i \omega ^3 r+\omega ^4\right) k^6\no\\
		&+36 r^5 \left(2448 r^5-972 i \omega  r^4-336 \omega ^2 r^3-1584 i \omega ^3 r^2-118 \omega ^4 r+33 i \omega ^5\right) k^4\no\\
		&+432 r^6 \omega ^2 \left(150 r^4-372 i \omega  r^3-147 \omega ^2 r^2-17 i \omega ^3 r-3 \omega ^4\right) k^2-1296 r^8 \omega ^4 \left(170 r^2+59 i \omega  r+2 \omega ^2\right)\bigg)\bigg)\no\\
		& -r^2 h(r)^2 \left(4 \left(k^2+6 r^2\right)-r^2 a_t'(r){}^2\right) \no\\
		&\times\bigg(3 k^4 r^8 a_t'(r)^{10}-4 k^2 r^6 \left(5 k^4+6 r (5 r-3 i \omega ) k^2+36 r^2 \omega ^2\right) a_t'(r)^8\no\\
		&+4 r^4 \left(4 k^8+48 r (r-i \omega ) k^6-6 r^2 \left(48 r^2+84 i \omega  r-55 \omega ^2\right) k^4+18 r^3 \omega ^2 (116 r+75 i \omega ) k^2-1071 r^4 \omega ^4\right) a_t'(r)^6\no\\
		&+16 r^2 a_t'(r)^4 \bigg(4 k^{10}+24 r (3 r-i \omega ) k^8+6 r^2 \left(192 r^2-60 i \omega  r+19 \omega ^2\right) k^6+54 r^3 \left(96 r^3-24 i \omega  r^2-66 \omega ^2 r+i \omega ^3\right) k^4\no\\
		&-9 r^4 \omega ^2 \left(1872 r^2+1044 i \omega  r-161 \omega ^2\right) k^2+270 r^5 \omega ^4 (57 r+10 i \omega )\bigg) -192 r^2 a_t'(r)^2\bigg(\left(48 r^2-72 i \omega  r-2 \omega ^2\right) k^8\no\\
		&+6 r \left(96 r^3-144 i \omega  r^2-56 \omega ^2 r+29 i \omega ^3\right) k^6+3 r^2 \left(576 r^4-864 i \omega  r^3-2520 \omega ^2 r^2+432 i \omega ^3 r-79 \omega ^4\right) k^4\no\\
		&-36 r^3 \omega ^2 \left(936 r^3-90 i \omega  r^2-179 \omega ^2 r+7 i \omega ^3\right) k^2+108 r^4 \omega ^4 \left(141 r^2+112 i \omega  r+\omega ^2\right)\bigg)\no\\
		& -2304 \bigg(2 \left(8 r^2-4 i \omega  r+\omega ^2\right) k^{10}+2 r \left(144 r^3-72 i \omega  r^2-2 \omega ^2 r+i \omega ^3\right) k^8\no\\
		&+r^2 \left(1728 r^4-864 i \omega  r^3+552 \omega ^2 r^2-492 i \omega ^3 r+25 \omega ^4\right) k^6+6 r^3 \bigg(576 r^5-288 i \omega  r^4+1464 \omega ^2 r^3\no\\
		&-828 i \omega ^3 r^2+19 \omega ^4 r+16 i \omega ^5\bigg) k^4+36 r^4 \omega ^2 \left(816 r^4-324 i \omega  r^3+19 \omega ^2 r^2-70 i \omega ^3 r-\omega ^4\right) k^2\no\\
		&+216 r^6 \omega ^4 \left(25 r^2-62 i \omega  r+11 \omega ^2\right)\bigg)\bigg) +2 r^4 \omega^2 h(r)\left(r^2 a_t'(r)^2-4 \left(k^2+6 r^2\right)\right)^2 \no\\
		&\times \bigg(2 k^2 r^6 a_t'(r)^8+r^4 \left(-8 k^4+6 r (i \omega -20 r) k^2+15 r^2 \omega ^2\right) a_t'(r)^6+6 r^3 \bigg(16 (2 r-i \omega ) k^4\no\\
		&+3 \left(96 r^3+11 \omega ^2 r\right) k^2+9 r^2 \omega^2 (5 i \omega -36 r)\bigg) a_t'(r)^4+48 r a_t'(r)^2 \bigg((8 r-2 i \omega ) k^6\no\\
		&+4 r \left(24 r^2+\omega ^2\right) k^4+9 r^2 \left(32 r^3-24 i \omega  r^2-2 \omega ^2 r-5 i \omega ^3\right) k^2+18 r^3 \omega ^2 \left(78 r^2-21 i \omega  r+4 \omega^2\right)\bigg)\no\\
		& -288 \bigg(\left(32 r^2-16 i \omega  r+\omega ^2\right) k^6+r \left(384 r^3-192 i \omega  r^2-8 \omega ^2 r+i \omega ^3\right) k^4\no\\
		&+36 r^3 \left(32 r^3-16 i \omega  r^2+9 \omega ^2 r-7 i \omega ^3\right) k^2+36 r^4 \omega^2 \left(68 r^2-27 i \omega  r+8 \omega ^2\right)\bigg)\bigg)\no\\
		& +r^7 \omega ^4 \left(a_t'(r)^2-24\right) \left(r a_t'(r)^2-24 r+12 i \omega \right) \left(r^2 a_t'(r)^2-4 \left(k^2+6 r^2\right)\right)^4\bigg\}\no\\
		&\hspace{-1cm}\mathcal{N}_h^2 =  - \frac{k a_t'(r)}{12 r h(r)} \bigg\{-995328 k^2 h(r)^6  \left(4 k^2+3 r^2 a_t'(r){}^2+8 i r \omega \right)\no\\
		& +41472 h(r)^5 \bigg(7 k^2 r^4 a_t'(r)^4+4 r^2 \left(10 k^4+r (78 r-7 i \omega ) k^2+42 r^2 \omega ^2\right) a_t'(r)^2\no
		\end{align}
		\begin{align}
		&+16 \left(3 k^6+r (18 r+5 i \omega ) k^4+6 r^2 \omega  (3 i r+2 \omega ) k^2-12 i r^3 \omega ^3\right)\bigg)\no\\
		 &-1728 h(r)^4 \bigg(5 k^2 r^6 a_t'(r)^6+12 r^4 \left(3 k^4+r (24 r+7 i \omega ) k^2-14 r^2 \omega ^2\right) a_t'(r)^4\no\\
		 &\hspace{-0.5cm}+16 r^2 \left(7 k^6+3 r (52 r-5 i \omega ) k^4+3 r^2 \left(228 r^2-8 i \omega  r+35 \omega ^2\right) k^2+6 r^3 \omega ^2 (150 r+i \omega )\right) a_t'(r)^2\no\\
		 &\hspace{-0.75cm}+192 \left(k^8+2 r (6 r+i \omega ) k^6+r^2 \left(36 r^2+2 i \omega  r+15 \omega ^2\right) k^4+2 r^3 \omega  \left(-30 i r^2+57 \omega  r-7 i \omega ^2\right) k^2+12 r^4 \omega ^3 (\omega -13 i r)\right)\bigg)\no\\
		 & +72 h(r)^3 \bigg(3 k^2 r^8 a_t'(r)^8+4 r^6 \left(2 k^4+r (65 i \omega -114 r) k^2+42 r^2 \omega ^2\right) a_t'(r)^6\no\\
		 &+16 r^4 \left(-4 k^6+r (6 r+7 i \omega ) k^4+6 r^2 \left(30 r^2+13 i \omega  r-12 \omega ^2\right) k^2+6 r^3 \omega ^2 (43 i \omega -228 r)\right) a_t'(r)^4\no\\
		 &+64 r^2 a_t'(r)^2 \bigg(-2 k^8+r (42 r-31 i \omega ) k^6+36 r^2 \left(20 r^2-4 i \omega  r+\omega ^2\right) k^4\no\\
		 &\hspace{-0.5cm}+6 r^3 \left(396 r^3+42 i \omega  r^2+198 \omega ^2 r-13 i \omega ^3\right) k^2+18 r^4 \omega ^2 \left(372 r^2+44 i \omega  r+3 \omega ^2\right)\bigg) \no\\
		 &+256 \bigg(k^{10}+r (18 r+7 i \omega ) k^8+6 r^2 \left(18 r^2+i \omega  r+7 \omega ^2\right) k^6+36 r^3 \left(6 r^3-19 i \omega  r^2+22 \omega ^2 r-i \omega ^3\right) k^4\no\\
		 &+18 r^4 \omega  \left(-156 i r^3+180 \omega  r^2-70 i \omega ^2 r+5 \omega ^3\right) k^2+108 r^6 \omega ^3 (9 \omega -58 i r)\bigg)\bigg)\no\\
		 &\hspace{-0.5cm} +6 r h(r)^2 \left(4 \left(k^2+6 r^2\right)-r^2 a_t'(r){}^2\right) \bigg(-3 k^2 r^7 a_t'(r)^8-2 r^5 \left(2 k^4+\left(3 i r \omega -60 r^2\right) k^2+54 r^2 \omega ^2\right) a_t'(r)^6\no\\
		 &+24 r^3 \left(2 k^6+8 r (r+2 i \omega ) k^4+r^2 \left(-24 r^2-18 i \omega  r+13 \omega ^2\right) k^2+2 r^3 \omega ^2 (144 r-13 i \omega )\right) a_t'(r)^4\no\\
		 &+32 r \bigg(2 k^8+3 r (4 r+3 i \omega ) k^6+24 r^2 \left(-3 r^2+3 i \omega  r+\omega ^2\right) k^4-12 r^3 \left(36 r^3-9 i \omega  r^2+45 \omega ^2 r-20 i \omega ^3\right) k^2\no\\
		 &-72 r^5 (45 r+8 i \omega ) \omega ^2\bigg) a_t'(r)^2-384 i \omega  \bigg(2 k^8+3 r (2 r-i \omega) k^6-6 r^2 \left(24 r^2+22 i \omega  r-\omega ^2\right)k^4\no\\
		 &-12 r^4 \left(54 r^2+57 i \omega  r+26 \omega ^2\right) k^2-72 r^5 (29 r+6 i \omega ) \omega ^2\bigg)\bigg) +i r \omega  h(r) \left(r^2 a_t'(r){}^2-4 \left(k^2+6 r^2\right)\right)^2 \no\\
		 &\times\bigg(-3 r^6 \left(k^2+2 i r \omega \right) a_t'(r)^6+4 r^4 \left(-k^4+6 r (5 r+3 i \omega ) k^2+3 r^2 \omega  (24 i r+5 \omega )\right) a_t'(r)^4\no\\
		 &\hspace{-0.5cm}+48 r^2 a_t'(r)^2\left(k^6+4 r (r+i \omega ) k^4-r^2 \left(12 r^2+24 i \omega  r+11 \omega ^2\right) k^2+3 r^3 \omega  \left(-24 i r^2+16 \omega  r+5 i \omega ^2\right)\right) \no\\
		 &+64 \bigg(k^8+6 r^2 k^6-18 r^2 \left(2 r^2+2 i \omega  r-\omega^2\right) k^4-9 r^3 \left(24 r^3+24 i \omega  r^2+14 \omega ^2 r-3 i \omega ^3\right) k^2\no\\
		 &\hspace{-0.5cm}-54 r^5 (26 r+5 i \omega ) \omega^2\bigg)\bigg) -i r^3 \omega^3 \left(r^2 a_t'(r)^2-4 \left(k^2+6 r^2\right)\right)^4 \left(r^2 a_t'(r)^2+4 \left(k^2-6 r^2\right)\right)\bigg\}\no\\
	\end{align}

\end{document}